\documentclass[preprint,prd,aps,showkeys,nofootinbib]{revtex4}
\usepackage{graphicx}

\usepackage{color}
\usepackage{CJK,upgreek,fancyhdr}
\usepackage{array}
\usepackage{booktabs}
\usepackage{multirow}

\UseRawInputEncoding

\begin{document}


\title{The leptonic di-flavor and di-number violation processes at high energy $\mu^\pm\mu^\pm $ colliders}

\author{Jin-Lei Yang$^{1,2}$\footnote{jlyang@hbu.edu.cn}, Chao-Hsi Chang$^{1,2,3}$\footnote{zhangzx@itp.ac.cn}, Tai-Fu Feng$^{4}$\footnote{fengtf@hbu.edu.cn}}
\affiliation{$^1$CAS Key Laboratory of Theoretical Physics, Institute of Theoretical Physics,
Chinese Academy of Sciences, Beijing 100190, China\\
$^2$School of Physical Sciences, University of Chinese Academy of Sciences, Beijing 100049, China\\
$^3$School of Physical Science and Technology, Lanzhou University, Lanzhou 730000, China\\
$^4$Department of Physics, Hebei University, Key Laboratory of High-precision Computation and Application of Quantum Field Theory of Hebei Province, Baoding, 071002, China}

\begin{abstract}
The leptonic di-flavor violation (LFV) processes $\mu^\pm \mu^\pm \rightarrow e^\pm e^\pm $, $\mu^\pm \mu^\pm \rightarrow \tau^\pm \tau^\pm $ and the leptonic di-number violation (LNV) processes $\mu^\pm \mu^\pm \rightarrow W^\pm _iW^\pm _j$ ($i,\;j=1,\;2$) at the same-sign high energy $\mu^\pm \mu^\pm $ colliders are studied. The new physics (NP) factors that may play roles in these processes are highlighted by cataloging them into three types. Taking into account the experimental constraints, the processes at $\mu^\pm\mu^\pm$ colliders are computed and the results are presented properly. The results lead to the conclusion that observing the NP factors through the LFV and LNV processes at TeV-energy $\mu^\pm\mu^\pm$ colliders has significant advantages that cannot be achieved elsewhere. Therefore, in developing the techniques of muon acceleration and collisions, the option of building the same-sign muon high-energy colliders should be considered seriously too.
\end{abstract}

\maketitle

\section{Introduction\label{sec1}}

Neutrino oscillations and flavor mixing among the three generations of neutrinos have been observed for several years~\cite{PDG}. These phenomena definitely mean the neutrinos have tiny but nonzero masses, and unambiguously are evidences of new physics (NP) beyond the standard model (SM). Regarding NP, one natural way for neutrinos to acquire tiny masses is through the so-called `seesaw mechanisms'~\cite{Schechter:1981bd}. In these mechanisms, neutrinos, along with newly introduced heavy neutral leptons, acquire Majorana components, allowing for leptonic di-flavor violation (LFV) processes $\mu^\pm \mu^\pm \rightarrow e^\pm e^\pm $, $\mu^\pm \mu^\pm \rightarrow\tau^\pm \tau^\pm $ and the leptonic di-number violation (LNV) processes $\mu^\pm \mu^\pm \rightarrow W^\pm W^\pm $ as well. Thus studying these kinds of the same-sign di-lepton and di-boson processes quantitatively to explore how the NP factors (such as the leptonic Majorana components, the right handed W-boson and the doubly charged Higgs etc) play roles is interesting and is the motivations for this paper. Furthermore, as it is known that flavor mixing parameters are strongly constrained by charged lepton flavor violation decays~\cite{MEG:2013oxv,SINDRUM:1987nra,BaBar:2009hkt,Hayasaka:2010np}, whereas the LFV and LNV processes considered in this study may not depend on the flavor mixing very much for the neutral leptons. Therefore, precisely observing the contributions of these mechanisms and the NP factors to these processes is particularly intriguing.

Considering the possibility of constructing high-energy $\mu^+ \mu^-$ colliders (e.g. $\sqrt s\geq 1.0$ TeV)\footnote{In comparison with electron, the essential difference for the high energy muon colliders from electron colliders is that the synchrotron radiation of muons in a circle ring is suppressed by a factor $(m_e/m_\mu)^4$}~\cite{Delahaye:2019omf,MICE:2019jkl,Eu:2019qin,Long:2020wfp} and so much important physics at the high energy colliders, we aim to investigate the important processes at high-energy same-sign $\mu^\pm \mu^\pm$ colliders~\cite{NeutrinoFactory:2002azy,Geer:2009zz,Heusch:1995yw,Shiltsev:2010qg}. If there is important physics present at high-energy same-sign $\mu^\pm \mu^\pm$ colliders comparable to that of high-energy $\mu^+\mu^-$ colliders, it is reasonable to consider both of them as potential future options. From the experiences for building a proton-antiproton ($p\bar{p}$) collider (Tevatron) and a proton-proton ($pp$) collider (LHC), there are no serious problems for building high energy same-sign $\mu^\pm \mu^\pm$ colliders in comparison with building high energy $\mu^+ \mu^-$ colliders, provided that the necessary techniques for muon sources, muon acceleration, and muon colliding are developed. Therefore, constructing high-energy same-sign $\mu^\pm \mu^\pm$ collider(s) is a natural extension of building high-energy $\mu^+ \mu^-$ colliders, as long as there is sufficient amount of interesting and significant physics to observe. Furthermore, the studies exploring the NP of same-sign $\mu^\pm \mu^\pm$ colliders are relatively fewer in the literatures compared to those focusing on $\mu^+ \mu^-$colliders~\cite{Chen:2016wkt,Buttazzo:2018qqp,Chiesa:2020awd,Han:2020uid,Han:2020pif,Han:2020uak,
Asadi:2021gah,Ruiz:2021tdt,Bottaro:2021snn}, so we would like quantitatively to investigate the characteristic processes, the LFV processes $\mu^\pm \mu^\pm \rightarrow e^\pm e^\pm $, $\mu^\pm \mu^\pm \rightarrow \tau^\pm \tau^\pm $ and the LNV processes $\mu^\pm \mu^\pm \rightarrow W^\pm _iW^\pm _j$ ($i,\;j=1,\;2$), at high-energy same-sign muon colliders in this work.

The process $e^\pm e^\pm \rightarrow \mu^\pm \mu^\pm $ was studied in Ref.~\cite{Cannoni:2003fh}, and the authors presented the theoretical predictions with the minimal type-I seesaw mechanism and analyzed the contributions from the supersymmetric particles. In Refs.~\cite{Raidal:1997tb,Cannoni:2002ny,Rodejohann:2010bv} the contributions from the doubly charged Higgs to the process $e^\pm e^\pm \to \mu^\pm \mu^\pm $ were also examined. The studies of the LNV di-boson process $e^\pm  e^\pm \to W_L^\pm  W_L^\pm $ were carried out in Refs.~\cite{Rizzo:1982kn,Dicus:1991fk,Belanger:1995nh,Ananthanarayan:1995cn,Gluza:1995ix,
Gluza:1995js,Greub:1996ct,Rodejohann:2010jh,Banerjee:2015gca,Asaka:2015oia,Wang:2016eln,Bandyopadhyay:2020mnp}. The theoretical predictions on the cross sections of $e^\pm e^\pm \to W_L^\pm  W_L^\pm $, $e^\pm e^\pm \rightarrow W_L^\pm  W_R^\pm $ in the left-right symmetric model (LRSM) were given in Refs.~\cite{London:1987nz,Gluza:1995ky,Helde:1994xj,Barry:2012ga}. In this work, we will investigate the NP contributions to the LFV di-lepton, LNV di-boson processes, and explore their phenomenological behaviors at the high energy $\mu^\pm \mu^\pm$ colliders. It is worth noting that the contributions to the LFV and LNV processes from $e$-flavor heavy neutral lepton are significantly constrained by the recent experimental upper bound on the $0\nu2\beta$ decay half-life, because the `core' process of the $0\nu2\beta$ decays is $d+d\rightarrow u+u+e+e$. Namely, all the processes with $e^\pm e^\pm $ being in the initial state are also constrained by the $0\nu2\beta$ experiments~\cite{Yang:2021ueh}.

In this work, we mainly focus on exploring the sources of NP which generate the LFV di-lepton and LNV di-boson processes, namely a). the neutral leptons' Majorana components with the help of the left-handed $W_L$ boson only or with the help of the left-handed and right-handed bosons $W_L$, $W_R$ both, and b). the doubly charged Higgs and the interference effects of the Higgs and the neutral leptons' Majorana components\footnote{The relevant supersymmetric (SUSY) particles can also make contributions to the considered processes, but the contributions are highly suppressed (significantly smaller compared to the contributions considered here) by their large masses and small flavor mixing parameters. Therefore, the present study will not consider the SUSY cases.}. To explore the phenomena of the NP factors and their combinations, we categorize them into three types: {\bf Type I of NP (TI-NP)} which, such as the $B-L$ symmetric SUSY model (B-LSSM)~\cite{Khalil:2008ps,Elsayed:2011de,Khalil:2015naa,Yang:2021duj}, involves the combination of neutral leptons' Majorana components and the left-handed $W_L$ boson; {\bf Type II of NP (TII-NP)} which, such as the left-right symmetric model (LRSM) without doubly charged Higgs, involves the combination of neutral leptons' Majorana components, the left-handed boson $W_L$, and the right-handed boson $W_R$; {\bf Type III of NP (TIII-NP}) which, such as the LRSM~\cite{Duka:1999uc,Dev:2016dja,Patra:2015bga,Mitra:2016kov,Maiezza:2016ybz,Mohapatra:2006gs,
Mohapatra:2005wg}, encompasses the presence of doubly charged Higgs alongside the $W_L$, $W_R$ bosons, and neutral leptons' Majorana components. The behaviors of the LFV di-lepton and LNV di-boson processes at high energy $\mu^\pm \mu^\pm$ colliders resulting from the three types of NP will be computed and discussed in this paper.

The paper is organized as follows: in Sec.~\ref{sec2} for later applications, the seesaw mechanisms, which give rise to the heavy neutral lepton masses, the tiny neutrino masses, the relevant Majorana components and the interaction vertices are outlined. In Sec.~\ref{sec3} the theoretical computations of the processes $\mu^\pm \mu^\pm \rightarrow l^\pm  l^\pm \;(l=e,\;\tau)$, $\mu^\pm \mu^\pm \rightarrow W^\pm _i W^\pm _j$ ($i,\;j=1,\;2$) for the three types of NP are given. In Sec.~\ref{sec4} the numerical results with suitable input parameters which are constrained by the available experiments are calculated and presented by figures. Finally, in Sec.~\ref{sec5} the results are discussed and a summary is made.

\section{The seesaw mechanisms and the relevant interactions\label{sec2}}

In this section, we briefly review the mechanisms which make the neutrinos to acquire tiny masses and mixtures. Furthermore, we outline the necessary interactions which relate to the three types of NP and are required for the computation of the LFV di-lepton and LNV di-boson processes.

Firstly, as a representative relating to {\bf TI-NP} let us consider the B-LSSM. Its gauge group is extended by adding a local group $U(1)_{B-L}$ to SM, where $B$, $L$ represent the baryon number and lepton number respectively. In this model, three right-handed neutral leptons and two singlet scalars (Higgs), possessing a non-zero $B-L$ charge, are introduced. The Majorana masses of the right-handed neutral leptons arise when the two singlet scalars (Higgs) acquire vacuum expectation values (VEVs).  Combining the Majorana mass terms with the Dirac mass terms, tiny neutrino masses can be obtained by the type-I seesaw mechanism. Namely the mass matrix of neutral leptons reads
\begin{eqnarray}
&&\left(\begin{array}{c}0,\;\;\;\;\;\;\;M_D^T\\M_D,\;\;\;\;M_R\end{array}\right),\label{eqa2}
\end{eqnarray}
where $M_D$ is the Dirac mass matrix of $3$ by $3$, and $M_R$ is the Majorana mass matrix of $3$ by $3$. We can define $\xi_{ij}=(M_D^TM_R^{-1})_{ij}$, then the mass matrix in Eq.~(\ref{eqa2}) can be diagonalized approximately as
\begin{eqnarray}
&&U_\nu^T\cdot\left(\begin{array}{c}0\qquad\xi M_R\\ M_R\xi^T\quad M_R\end{array}\right)\cdot U_\nu\approx\left(\begin{array}{c}\hat m_\nu\qquad0\\ 0\qquad \hat M_N\end{array}\right),
\end{eqnarray}
where $\hat m_\nu={\rm diag}(m_{\nu_1},m_{\nu_2},m_{\nu_3})$ with $m_{\nu_i}\;(i=1,\;2,\;3)$ denoting the $i-$th generation of light neutrino masses, $\hat M_N={\rm diag}(M_{N_1},M_{N_2},M_{N_3})$ with $M_{N_i}\;(i=1,\;2,\;3)$ denoting the $i-$th generation of heavy neutral lepton masses, and
\begin{eqnarray}
&&U_\nu\equiv\left(\begin{array}{c}U\quad S\\T\quad V\end{array}\right)=\left(\begin{array}{c}1-\frac{1}{2}\xi^*\xi^T\qquad\xi^*\\ -\xi^T\qquad1-\frac{1}{2}\xi^T\xi^*\end{array}\right)\cdot\left(\begin{array}{c}U_{L}\qquad0\\0\qquad U_R\end{array}\right).\label{eqa3}
\end{eqnarray}
Then we have
\begin{eqnarray}
&&U_L\cdot \hat m_\nu\cdot U_L^T=-M_D^T M_R^{-1}M_D,\nonumber\\
&&U_R\cdot \hat M_N\cdot U_R^T=M_R.\label{eq4}
\end{eqnarray}
In the following analysis, the $3$ by $3$ matrix $U$ is taken as the Pontecorvo-Maki-Nakagawa-Sakata (PMNS) mixing matrix which is measured from the neutrino oscillation experiments. The relevant Lagrangian for the $l-W-\nu$ and $l-W-N$ ($l=e,\mu,\tau$) interactions in the model becomes
\begin{eqnarray}
&&\mathcal{L}_{W}^{BL}=\frac{ig_2}{\sqrt2}\sum_{j=1}^3\Big[U_{ij}\bar l_i\gamma^\mu P_L\nu_{j}W_{L,\mu}^-+S_{ij}\bar l_i\gamma^\mu P_LN_{j}W_{L,\mu}^-+h.c\Big],\label{eqa5}
\end{eqnarray}
where $P_{L,R}=(1\mp\gamma^5)/2$, and $\nu,\;N$ are the four-component forms of mass eigenstates corresponding to light, heavy neutral leptons respectively. The LFV di-lepton processes will be calculated in the Feynman gauge later on, hence the $l-G-\nu$ and $l-G-N$ interactions where $G$ denotes the Goldstone boson will occur, and the relevant Lagrangian is written as
\begin{eqnarray}
&&\mathcal{L}_{G}^{BL}=\frac{ig_2}{\sqrt2 M_{W_L}}\sum_{j=1}^3\Big\{\bar l_i\Big[(M_D^\dagger\cdot T^*)_{ij} P_R-(\hat m_l\cdot U)_{ij} P_L\Big]\nu_{j}G_{L}^-\nonumber\\
&&\qquad\qquad\qquad\qquad +\bar l_i\Big[(M_D^\dagger\cdot V^*)_{ij} P_R-(\hat m_l\cdot S)_{ij} P_L\Big]N_{j}G_{L}^-+h.c\Big\},\label{eq6}
\end{eqnarray}
where $U,\;S,\;T,\;V$ are matrices of $3$ by $3$ defined in Eq.~(\ref{eqa3}), $\hat m_l={\rm diag}(m_{e},m_{\mu},m_{\tau})$ with $m_{e},m_{\mu},m_{\tau}$ denoting the charged lepton masses, $G_L$ being the goldstone boson which is ``eaten" by $W_L$ in unitary gauge, $M_{W_L}$ is the $W_L$ boson mass. Having the relevant lagrangian $\mathcal{L}_{W}^{BL}$ and $\mathcal{L}_{G}^{BL}$, the theoretical predictions of the considered processes for {\bf TI-NP} can be calculated.

As a representative relating to the  {\bf TII-NP} and {\bf TIII-NP}, let us consider the left-right symmetric model (LRSM). Its gauge group is $SU(3)_C\bigotimes SU(2)_L\bigotimes SU(2)_R\bigotimes U(1)_{B-L}$. The additional three generations of the right-handed neutral leptons which with the three generations of the right-handed charged leptons form doublets, the di-doublet scalar (Higgs) and the two triplet scalars (Higgs)
\begin{eqnarray}
&&\psi_R=\left(\begin{array}{c}N_R\\l_R\end{array}\right),\;\Phi=\left(\begin{array}{c}\phi_1^0\quad \phi_2^+\\ \phi_1^-\quad \phi_2^0\end{array}\right),\;\Delta_{L,R}=\left(\begin{array}{c}\Delta_{L,R}^+/\sqrt 2\quad \Delta_{L,R}^{++}\\ \Delta_{L,R}^0\quad -\Delta_{L,R}^+/\sqrt 2\end{array}\right)
\end{eqnarray}
are involved and $v_1,\;v_2,\;v_L,\;v_R$ are the VEVs of $\phi_1^0,\;\phi_2^0,\;\Delta_{L}^0,\;\Delta_{R}^0$ respectively. The Yukawa Lagrangian for the lepton sector is given by
\begin{eqnarray}
&&\mathcal{L}_Y=-h_{ij}\psi_{L,i}^\dagger\Phi\psi_{R,j}-\tilde h_{ij}\psi_{L,i}^\dagger\tilde\Phi\psi_{R,j}-Y_{L,ij}\psi_{L,i}^TC(-i\sigma^2)\Delta_L\psi_{L,j}\nonumber\\
&&\qquad\;\;-Y_{R,ij}\psi_{R,i}^TC(i\sigma^2)\Delta_R\psi_{R,j}+h.c.,\label{eq88}
\end{eqnarray}
where the family indices $i,\;j$ are summed over, $\tilde \Phi=\sigma^2\Phi^*\sigma^2$. The tiny neutrino masses are obtained by both of the type-I and type-II seesaw mechanisms when the Higgs $\phi_1^0,\;\phi_2^0,\;\Delta_{L}^0,\;\Delta_{R}^0$ achieve VEVs~\cite{Duka:1999uc,Dev:2016dja,Patra:2015bga,Mitra:2016kov,Maiezza:2016ybz}. Then the mass matrix of neutral leptons can be written as
\begin{eqnarray}
&&\left(\begin{array}{c}M_L,\;\;\;\;M_D^T\\M_D,\;\;\;\;M_R\end{array}\right),\label{eq7}
\end{eqnarray}
where
\begin{eqnarray}
&&M_D=\frac{1}{\sqrt 2}(h v_1+\tilde h v_2)^T,\;M_L=\sqrt 2 Y_L v_L,\;M_R=\sqrt 2 Y_R v_R.\label{eq71}
\end{eqnarray}
The Lagrangian for the $l^--\Delta^{--}_{L,R}-l^-$ interactions is
\begin{eqnarray}
&&\mathcal{L}_{\Delta ll}^{LR}=i2Y_{L,ij}\bar l_iP_Ll_j^C\Delta_L^{--}+i2Y_{R,ij}\bar l_iP_Rl_j^C\Delta_R^{--}.\label{eq72}
\end{eqnarray}

The mass matrix in Eq.~(\ref{eq7}) can be diagonalized in terms of a unitary matrix $U_\nu$, whereas the matrix $U_\nu$ can be expressed similarly as that in the above case of the B-LSSM Eq.~(\ref{eqa3}). Then we can obtain~\cite{Yang:2021ueh}
\begin{eqnarray}
&&U_L\cdot \hat m_\nu\cdot U_L^T\approx M_L-M_D^T M_R^{-1}M_D,\nonumber\\
&&U_R\cdot \hat M_N\cdot U_R^T\approx M_R+\frac{1}{2}M_R^{-1}M_D^*M_D^T+\frac{1}{2}M_DM_D^\dagger M_R^{-1}\approx M_R.   \label{eq8}
\end{eqnarray}
The second formula in Eq.~(\ref{eq8}) are obtained by using the approximation $M_R^{-1}M_D^*M_D^T\approx M_DM_D^\dagger M_R^{-1}\ll M_R$. Due to the $SU(2)_R$ gauge group, there are charged right-handed gauge bosons $W^\pm_R$ additionally and the two kinds of bosons $W_L$ and $W_R$ may be mixed. As the result of $W_L-W_R$ mixing, the masses of physical $W_1$ (dominated by the left-handed $W_L$) and $W_2$ (dominated by the right-handed $W_R$) can be written as~\cite{Yang:2021ueh}:
\begin{eqnarray}
&&M_{W_1}\approx\frac{g_2}{2}(v_1^2+v_2^2)^{1/2}, \;\;\;\;\; M_{W_2}\approx\frac{g_2}{\sqrt2}v_R,\label{eq9}
\end{eqnarray}
Then the Lagrangian for the $l-W-\nu$ and $l-W-N$ interactions for the LRSM is
\begin{eqnarray}
&&\mathcal{L}_W^{LR}=\frac{ig_2}{\sqrt2}\sum_{j=1}^3\Big[\bar l_i(\cos\zeta U_{ij}\gamma^\mu P_L+\sin\zeta T^*_{ij}\gamma^\mu P_R)\nu_{j}W_{1,\mu}^-\nonumber\\
&&\qquad\qquad\;\;\;+\bar l_i(\cos\zeta T^*_{ij}\gamma^\mu P_R-\sin\zeta U_{ij}\gamma^\mu P_L)\nu_{j}W_{2,\mu}^-\nonumber\\
&&\qquad\qquad\;\;\;+\bar l_i(\cos\zeta S_{ij}\gamma^\mu P_L+\sin\zeta V^*_{ij}\gamma^\mu P_R)N_{j}W_{1,\mu}^-\nonumber\\
&&\qquad\qquad\;\;\;+\bar l_i(\cos\zeta V^*_{ij}\gamma^\mu P_R-\sin\zeta S_{ij}\gamma^\mu P_L)N_{j}W_{2,\mu}^-+h.c\Big],\label{eq10}
\end{eqnarray}
where $\tan2\zeta=\frac{2v_1v_2}{v_R^2-v_L^2}$ denotes the mixing between $W_L$ and $W_R$, the matrices $U,\;S,\;T,\;V$ are defined in Eq.~(\ref{eqa3}), $\nu,\;N$ are the four-component fermion fields of the mass eigenstates corresponding
to light and heavy neutral leptons respectively. The Lagrangian for the $l-G-\nu$ and $l-G-N$ interactions may be written as~\cite{Duka:1999uc}
\begin{eqnarray}
&&\mathcal{L}_G^{LR}=\frac{ig_2}{\sqrt2 M_{W_L}}\sum_{j=1}^3\Big[\bar l_i(\lambda_{1,ij} P_L+\lambda_{2,ij} P_R)\nu_jG_{L}^-+\bar l_i(\lambda_{3,ij} P_L+\lambda_{4,ij} P_R)N_{j}G_{L}^-\nonumber\\
&&\qquad\qquad\qquad\qquad +\bar l_i(\lambda_{5,ij} P_L)\nu_jG_{R}^-+\bar l_i(\lambda_{6,ij} P_L)N_{j}G_{R}^-+h.c.\Big],\label{eq11}
\end{eqnarray}
where $G_L, G_R$ are the unphysical Goldstone bosons when the Feynman gauge is applied \footnote{For convenience we apply the Feynman gauge to computing the processes in this paper.}, and
\begin{eqnarray}
&&\lambda_1=-\hat m_l^\dagger\cdot U,\qquad\;\;\lambda_2=M_D^\dagger\cdot T^*,\qquad\;\;\lambda_3=-\hat m_l^\dagger\cdot S\nonumber\\
&&\lambda_4=M_D^\dagger\cdot V^*,\qquad\;\;\lambda_5=\frac{M_{W_1}}{M_{W_2}}M_R^\dagger\cdot T,\quad\lambda_6=\frac{M_{W_1}}{M_{W_2}}M_R^\dagger\cdot V^*.\label{eq12}
\end{eqnarray}
Finally, the relevant Lagrangian for the $W-\Delta^{--}_{L,R}-W$ interactions is
\begin{eqnarray}
&&\mathcal{L}_{\Delta WW}^{LR}=i\sqrt2 g_2^2 v_L\Delta_L^{--}W_1^{\mu+}W_{1\mu}^++i\sqrt2 g_2^2 v_L\sin\zeta\Delta_L^{--}W_1^{\mu+}W_{2\mu}^+\nonumber\\
&&\qquad\qquad +i\sqrt2 g_2^2 v_R\sin\zeta\Delta_R^{--}W_1^{\mu+}W_{2\mu}^++i\sqrt2 g_2^2 v_R\Delta_R^{--}W_2^{\mu+}W_{2\mu}^++h.c..
\end{eqnarray}
Hence now, the theoretical predictions of the considered processes for {\bf TII-NP} can be calculated based on the Lagrangian $\mathcal{L}_{W}^{LR}$, $\mathcal{L}_{G}^{LR}$, and the ones for {\bf TIII-NP} can be calculated based on the Lagrangian $\mathcal{L}_{\Delta ll}^{LR}$, $\mathcal{L}_{W}^{LR}$, $\mathcal{L}_{G}^{LR}$, $\mathcal{L}_{\Delta WW}^{LR}$.

Note that there is a possible model which contains doubly charged Higgs in its scalar triplet with Type-II seesaw~\cite{Mohapatra:1979ia,Schechter:1980gr,Cheng:1980qt,Lazarides:1980nt}, and the doubly charged component of the triplet Higgs can also cause the LFV and LNV processes. But the contributions to the processes are highly suppressed (which are much smaller than those considered here) as the relevant interactions are proportional to the light neutrino masses. Thus the contributions to the LFV and LNV processes from the three types of NP considered here are much great that we will not discuss the model here.

\section{Computations of the processes $\mu^\pm \mu^\pm \rightarrow l^\pm  l^\pm \;(l=e,\;\tau)$ and $\mu^\pm \mu^\pm \rightarrow W^\pm _i W^\pm _j$ ($i,j=1,2$)\label{sec3}}

In this section we will provide the calculations of the processes $\mu^\pm \mu^\pm \rightarrow l^\pm  l^\pm \;(j=e,\;\tau)$, $\mu^\pm \mu^\pm \rightarrow W^\pm _i W^\pm _j\;(i,j=1,2)$ within {\bf TI-NP}, {\bf TII-NP} and {\bf TIII-NP}. The results for {\bf TI-NP} and {\bf TII-NP} can be obtained by switching off certain interactions from the relevant results for {\bf TIII-NP}. In all computations, the flavor mixing parameters of the heavy neutral leptons are ignored, because the flavor mixing parameters are constrained strongly by the charged lepton flavor violating decays~\cite{MEG:2013oxv,SINDRUM:1987nra,BaBar:2009hkt,Hayasaka:2010np}.

\subsection{The LFV processes $\mu^\pm \mu^\pm \rightarrow l^\pm  l^\pm \;(l=e,\;\tau)$\label{sec3.1}}

\begin{figure}
\setlength{\unitlength}{1mm}
\centering
\includegraphics[width=5in]{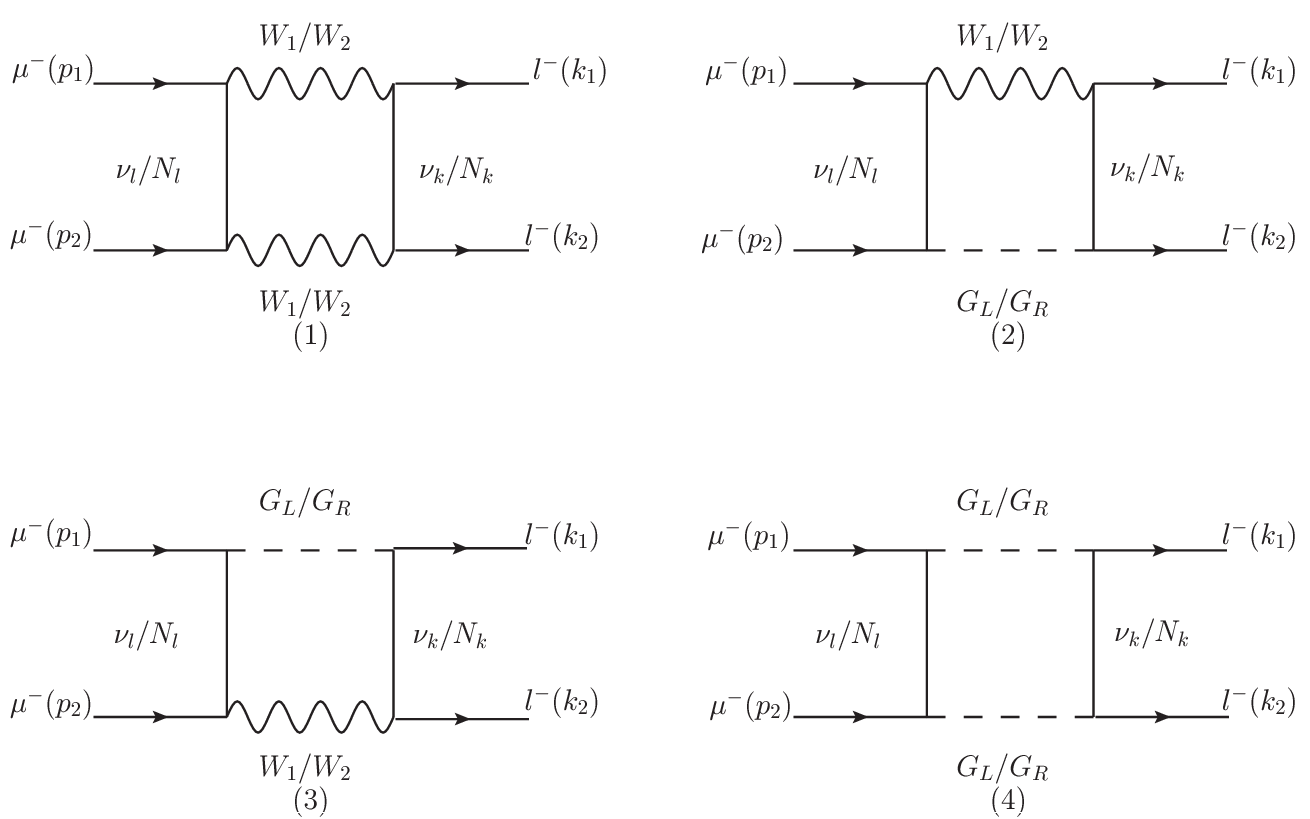}
\vspace{0cm}
\caption[]{The Feynman diagrams (leading order) in Feynman gauge for the $\mu^- \mu^- \rightarrow l^-  l^- \;(l=e,\;\tau)$ processes via neutrino and neutral heavy lepton exchanges.}
\label{Fllll}
\end{figure}
The leading-order Feynman diagrams for the LFV di-lepton processes $\mu^\pm \mu^\pm \rightarrow l^\pm  l^\pm \;(l=e,\;\tau)$, incorporating the contributions from Majorana neutral leptons in {\bf TI-NP}, {\bf TII-NP} and {\bf TIII-NP} are collected in Fig.~\ref{Fllll}, but in the case of {\bf TI-NP}, the Feynman diagrams with $W_2$ and/or $G_R$ line(s) should be moved away. Note that, here for simplification, the Feynman diagrams with the gauge boson $W_{1,2}$ lines and/or the Goldstone boson $G_{L,R}$ lines being crossed are not presented in the figure, in fact the contributions from these crossing diagrams to the processes should be well taken into account. Hence when doing the calculations, we do consider the contributions from these crossing diagrams.

The mixing parameter $\zeta$ between the left-handed boson $W_L$ and the right-handed boson $W_R$ is constrained in the range $\zeta\lesssim7.7\times10^{-4}$~\cite{Dev:2014xea}, and the charged lepton masses are much smaller than the center-of-mass energy of the collisions. So in the calculations we ignore the contributions proportional to $\mathcal{O}(\zeta^2)$ and charged lepton masses. Then the amplitude of the processes $\mu^\pm \mu^\pm \rightarrow l^\pm  l^\pm $ for {\bf TIII-NP} can be formulated as
\begin{eqnarray}
&&\mathcal{M}(\mu^\pm \mu^\pm \rightarrow l^\pm  l^\pm )\approx \frac{i}{16\pi^2}\sum_{X,Y=L,R}\Big(C_1^{XY} \mathcal{O}_1^{XY}+C_2^{XY} \mathcal{O}_2^{XY}+C_3^{XY} \mathcal{O}_3^{XY}\nonumber\\
&&\qquad\qquad\qquad\qquad +C_4^{XY} \mathcal{O}_4^{XY}+C_5^{XY} \mathcal{O}_5^{XY}+C_6^{XY} \mathcal{O}_6^{XY}+C_7^{XY} \mathcal{O}_7^{XY}\Big),\label{eq13}
\end{eqnarray}
where
\begin{eqnarray}
&&\mathcal{O}_1^{XY}=\bar u(k_1)\gamma_{\mu}P_X u^c(k_2)\overline{u^c}(p_2)\gamma^{\mu}P_Y u(p_1),\nonumber\\
&&\mathcal{O}_2^{XY}=\bar u(k_1)p\!\!\!/_1 P_X u^c(k_2)\overline{u^c}(p_2)k\!\!\!/_1P_Y u(p_1),\nonumber\\
&&\mathcal{O}_3^{XY}=\bar u(k_1)P_Xu^c(k_2)\overline{u^c}(p_2)P_Y u(p_1),\nonumber\\
&&\mathcal{O}_4^{XY}=\bar u(k_1)\gamma_\mu P_Xu^c(k_2)\overline{u^c}(p_2)\gamma^\mu k\!\!\!/_1P_Y u(p_1),\nonumber\\
&&\mathcal{O}_5^{XY}=\bar u(k_1)\gamma_\mu p\!\!\!/_1 P_Xu^c(k_2)\overline{u^c}(p_2)\gamma^\mu P_Y u(p_1),\nonumber\\
&&\mathcal{O}_6^{XY}=\bar u(k_1)P_Xu^c(k_2)\overline{u^c}(p_2)k\!\!\!/_1P_Y u(p_1),\nonumber\\
&&\mathcal{O}_7^{XY}=\bar u(k_1)p\!\!\!/_1 P_Xu^c(k_2)\overline{u^c}(p_2)P_Y u(p_1),\label{eq14}
\end{eqnarray}
where $u$ is the Dirac spinor of the leptons, $u^c\equiv C\bar u^T$ is its charge conjugation, the charge conjugation operator $C\equiv i\gamma_2\gamma_0$, and $\gamma_{0},\;\gamma_{2}$ are the Dirac matrices. The momenta $p_1,p_2,k_1,k_2$ are defined as shown in Fig.~\ref{Fllll}, the coefficients $C_i^{XY}$ in Eq.~(\ref{eq13}) can be read out from the amplitudes relating to the Feynman diagrams Fig.~\ref{Fllll}.

To show how the coefficients are read out, now let us take the Feynman diagram Fig.~\ref{Fllll} (1) as an example, where $\nu_{k}$, $\nu_{l}$, $W_1$, $W_2$ appear in the loop. Firstly, according to the interactions in Sec.~\ref{sec2}, the amplitude can be written as
\begin{eqnarray}
&&\mathcal{M}(\nu\nu W_1W_2)=\frac{1}{4}g_2^4\mu^{4-D}\int \frac{d^Dk}{(2\pi)^D}\bar u(k_1)(\cos\zeta U_{jk}\gamma_\mu P_L+\sin\zeta T^*_{jk}\gamma_\mu P_R)(k\!\!\!/+k\!\!\!/_1-p\!\!\!/_1\nonumber\\
&&\qquad +m_{\nu_k})(\cos\zeta T_{jk}^*\gamma_\nu P_L-\sin\zeta U_{jk}\gamma_\nu P_R)u^c(k_2)\overline{u^c}(p_2) (\cos\zeta T_{2l}\gamma^\nu P_L\nonumber\\
&&\qquad -\sin\zeta U^*_{2l}\gamma^\nu P_R)(k\!\!\!/+m_{\nu_l})(\cos\zeta U_{2l}^*\gamma^\mu P_L+\sin\zeta T_{2l}\gamma^\mu P_R)u(p_1)\frac{1}{k^2-m_{\nu_l}^2}\nonumber\\
&&\qquad \frac{1}{(k+k_1-p_1)^2-m_{\nu_i}^2}\frac{1}{(k-p_1)^2-M_{W_1}^2+i\Gamma_{W_1} M_{W_1}}\frac{1}{(k+p_2)^2-M_{W_2}^2+i\Gamma_{W_2} M_{W_2}},\label{a3}\nonumber\\
\end{eqnarray}
where $\Gamma_{W_1},\;\Gamma_{W_2}$ are the total decay widthes of $W_1,\;W_2$ bosons respectively.
The integrals appearing in Eq.~(\ref{a3}) can be calculated by using Loop-Tools~\cite{Denner:1991kt,Hahn:1998yk}, hence for the further calculations, we define the functions following the conventions in Loop-Tools as
\begin{eqnarray}
&&D_0\equiv\frac{\mu^{4-D}}{i\pi^{D/2}r_\Gamma}\int\frac{d^D q}{[q^2-m_1^2][(q+l_1)^2-m_2^2]
[(q+l_2)^2-m_3^2][(q+l_3)^2-m_4^2]},\nonumber\\
&&D_\mu\equiv\frac{\mu^{4-D}}{i\pi^{D/2}r_\Gamma}\int\frac{d^D q\;\;q_\mu}{[q^2-m_1^2][(q+l_1)^2-m_2^2]
[(q+l_2)^2-m_3^2][(q+l_3)^2-m_4^2]}\nonumber\\
&&\quad\;\;=\sum_{i=1}^3 l_{i\mu}D_i,\nonumber\\
&&D_{\mu\nu}\equiv\frac{\mu^{4-D}}{i\pi^{D/2}r_\Gamma}\int\frac{d^D q\;\;q_\mu q_\nu}{[q^2-m_1^2][(q+l_1)^2-m_2^2]
[(q+l_2)^2-m_3^2][(q+l_3)^2-m_4^2]}\nonumber\\
&&\quad\;\;\;=g_{\mu\nu}D_{00}+\sum_{i,j=1}^3 l_{i\mu}l_{j\nu}D_{ij}\,,\label{a4}
\end{eqnarray}
where $l_1,\;l_2,\;l_3$ are combinations of out-leg particles' momentum (for example, for the loop integral in Eq.~(\ref{a3}), we may have $l_1=k_1-p_1,\;l_2=-p_1,\;l_3=p_2$), $q$ is the integral momentum, $m_i^2\equiv M_i^2-i\Gamma_i M_i$ with $M_i$, $\Gamma_i$ denoting the mass and total decay width of loop particle $i$ respectively, and
$$r_\Gamma=\frac{\Gamma^2(1-\varepsilon)\Gamma(1+\varepsilon)}{\Gamma(1-2\varepsilon)},\;\;D=4-2\varepsilon$$

According to the definitions in Eq.~(\ref{a4}), the amplitude $\mathcal{M}(\nu\nu W_1W_2)$ can be simplified by neglecting the tiny neutrino mass terms in the numerators of the neutrino propagators and the terms proportional to charged lepton masses which appear after applying the on-shell condition for the leptons. Then the amplitude $\mathcal{M}(\nu\nu W_1W_2)$ becomes
\begin{eqnarray}
&&\mathcal{M}(\nu\nu W_1W_2)\approx\frac{i}{64\pi^2}g_2^4\cos^4\zeta U_{jk}T^*_{jk}U^*_{2l}T_{2l}\Big[4D_{00}\bar u(k_1)\gamma_{\mu}P_L u^c(k_2)\overline{u^c}(p_2)\gamma^{\mu}P_L u(p_1)\nonumber\\
&&\qquad\qquad\qquad\;\;\; +4(D_0+D_1+D_2+D_{12})\bar u(k_1)p\!\!\!/_1 P_L u^c(k_2)\overline{u^c}(p_2)k\!\!\!/_1P_L u(p_1)\Big],
\end{eqnarray}
where $D_{00},\;D_{0},\;D_{1},\;D_{2},\;D_{12}$ can be computed numerically by using Loop-Tools. Then from $\mathcal{M}(\nu\nu W_1W_2)$, the coefficients $C_i^{XY}$ of $\mathcal{O}_i^{XY}\;(i=1,...,7)$, defined in Eq.~(\ref{eq14}), can be read out as
\begin{eqnarray}
&&C_1^{LL}(\nu\nu W_1W_2)=\frac{1}{4}g_2^4\cos^4\zeta U_{jk}T^*_{jk}U^*_{2l}T_{2l}(4D_{00}),\nonumber\\
&&C_2^{LL}(\nu\nu W_1W_2)=\frac{1}{4}g_2^4\cos^4\zeta U_{jk}T^*_{jk}U^*_{2l}T_{2l}[4(D_0+D_1+D_2+D_{12})]\,.
\end{eqnarray}
Namely corresponding to the amplitude for the Feynman diagram Fig.~\ref{Fllll} (1), only the coefficients $C_1^{LL}$ and $C_2^{LL}$ receive nonzero contributions. When we calculate the other Feynman diagrams of Fig.~\ref{Fllll} where the two outgoing charged leptons $l$ ($l=e,\;\tau$) of the Feynman diagrams are alternated, and the identity formulas
\begin{eqnarray}
&&[\bar u(k_2)\gamma_{\mu}P_X u^c(k_1)]^T=\bar u(k_1)\gamma_{\mu}P_Y u^c(k_2)\,,\nonumber\\
&&[\bar u(k_2)P_X u^c(k_1)]^T=-\bar u(k_1)P_X u^c(k_2). \label{eq15}
\end{eqnarray}
are applied to $\mathcal{O}_i^{XY}\,(i=1,...,7)$ defined in Eq.~(\ref{eq14}). Because each amplitude of the above Feynman diagrams may be treated in the same way as Fig.~\ref{Fllll} (1), so all the coefficients $C_i^{XY}$ in Eq.~(\ref{eq13}) for the relevant processes may be obtained.

\begin{figure}
\setlength{\unitlength}{1mm}
\centering
\includegraphics[width=2in]{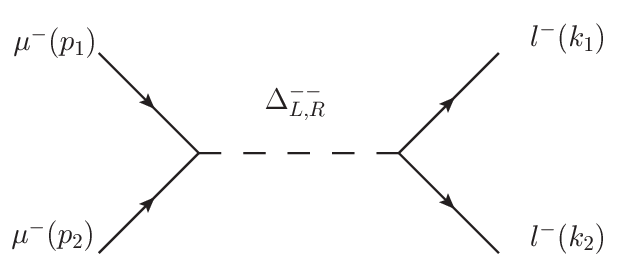}
\vspace{0cm}
\caption[]{The Feynman diagrams for $\mu^\pm \mu^\pm \rightarrow l^\pm  l^\pm \;(l=e,\;\tau)$ due to doubly charged Higgs.}
\label{FllllDC}
\end{figure}

The contributions from doubly charged Higgs ($\Delta^{\pm \pm}$) to the LFV di-lepton processes $\mu^\pm \mu^\pm \rightarrow l^\pm  l^\pm \;(l=e,\;\tau)$ can be estimated from the Feynman diagram in Fig.~\ref{FllllDC}.
The amplitude corresponding to Fig.~\ref{FllllDC} can be formulated into that as Eq.~(\ref{eq13}), and the nonzero
coefficients are
\begin{eqnarray}
&&C_3^{LR}(\Delta_L^{--})=\frac{4Y_{L,22}Y_{L,jj}}{(p_1+p_2)^2-M_{\Delta_L^{--}}^2+
i\Gamma_{\Delta_L^{--}}M_{\Delta_L^{--}}},\nonumber\\
&&C_3^{RL}(\Delta_R^{--})=\frac{4Y_{R,22}Y_{R,jj}}{(p_1+p_2)^2-M_{\Delta_R^{--}}^2+
i\Gamma_{\Delta_R^{--}}M_{\Delta_L^{--}}},
\end{eqnarray}
are left, where $j=e,\;\tau$, $\Gamma_{\Delta_{L,R}^{--}}$ is the total decay width of the Higgs $\Delta_{L,R}^{--}$.

Based on the total amplitude formulated as Eq.~(\ref{eq13}), the amplitude can be squared absolutely by summing up the lepton spins in the initial and final states. Neglecting the charged lepton masses which is turned out from the square of $\mathcal{O}_i^{XY}\, (i=1,...,7)$, those in Eq.~(\ref{eq14}), the result can be simplified as
\begin{eqnarray}
&&|\mathcal{M}(\mu^\pm \mu^\pm \rightarrow l^\pm  l^\pm )|^2\approx \frac{1}{64\pi^4}\Big\{4(p_1\cdot k_2)^2(|C_1^{LL}|^2+|C_1^{RR}|^2)+4(p_1\cdot k_1)^2(|C_1^{LR}|^2\nonumber\\
&&\qquad\qquad\quad\;\;+|C_1^{RL}|^2)+4(p_1\cdot k_1)^2(p_1\cdot k_2)^2(|C_2^{LL}|^2+|C_2^{RR}|^2+|C_2^{LR}|^2\nonumber\\
&&\qquad\qquad\quad\;\;+|C_2^{RL}|^2)+(p_1\cdot p_2)^2(|C_3^{LL}|^2+|C_3^{RR}|^2+|C_3^{LR}|^2+|C_3^{RL}|^2)\nonumber\\
&&\qquad\qquad\quad\;\;+8p_1\cdot k_1(p_1\cdot k_2)^2{\rm Re}[C_1^{LL}C_2^{LL*}+C_1^{RR}C_2^{RR*}]+4p_1\cdot k_1[(p_1\cdot k_1)^2\nonumber\\
&&\qquad\qquad\quad\;\;+(p_1\cdot k_2)^2-(p_1\cdot p_2)^2]{\rm Re}[C_1^{LR}C_2^{LR*}+C_1^{RL}C_2^{RL*}]\nonumber\\
&&\qquad\qquad\quad\;\; +2p_1\cdot k_1 p_1\cdot k_2 p_1\cdot p_2[4(|C_4^{LR}|^2+|C_4^{RL}|^2+|C_5^{LL}|^2+|C_5^{RR}|^2)\nonumber\\
&&\qquad\qquad\quad\;\;+|C_6^{LL}|^2+|C_6^{RR}|^2+|C_6^{LR}|^2+|C_6^{RL}|^2+|C_7^{LL}|^2+|C_7^{RR}|^2+
|C_7^{LR}|^2\nonumber\\
&&\qquad\qquad\quad\;\;+|C_7^{RL}|^2]+4p_1\cdot k_1[(p_1\cdot p_2)^2+(p_1\cdot k_2)^2-(p_1\cdot k_1)^2]{\rm Re}[C_5^{LL}C_6^{LL*}\nonumber\\
&&\qquad\qquad\quad\;\;+C_5^{RR}C_6^{RR*}-C_4^{LR}C_7^{LR*}-C_4^{RL}C_7^{RL*}]\Big\}.\label{eq16}
\end{eqnarray}
Now the cross section of LFV di-lepton processes can be written as
\begin{eqnarray}
&&\sigma=\frac{1}{64\pi s}\int_{-1}^1\frac{1}{4}|\mathcal{M}(\mu^\pm\mu^\pm\rightarrow l^\pm l^\pm )|^2 {\rm d}\cos\theta.
\end{eqnarray}
where the factor $\frac{1}{4}$ comes from averaging the lepton spins in the initial state, $\theta$ is the angle between the direction of the outgoing lepton $l$ with the collision axis, $\sqrt s$ is the total collision energy of $\mu^\pm \mu^\pm$ colliders.

\subsection{The LNV processes $\mu^\pm \mu^\pm \rightarrow  W_i^\pm W_j^\pm (i,j=1,2)$\label{sec3.2}}

\begin{figure}
\setlength{\unitlength}{1mm}
\centering
\includegraphics[width=5in]{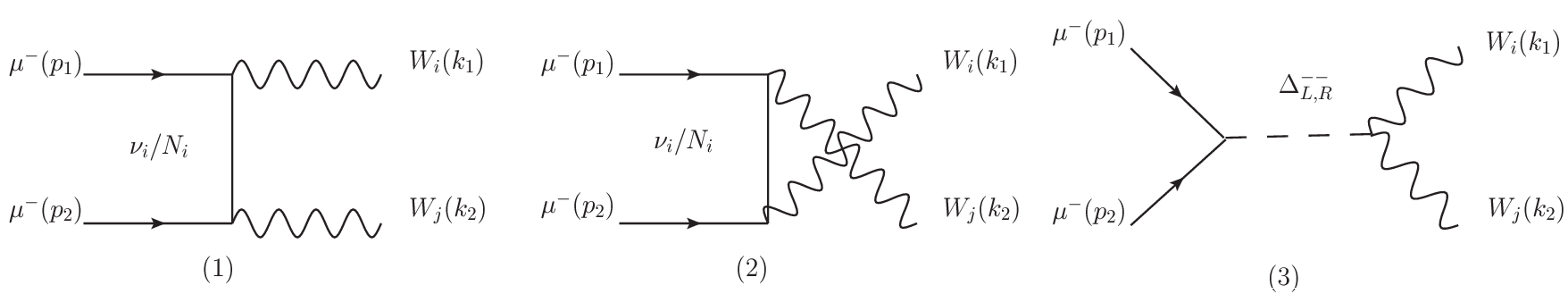}
\vspace{0cm}
\caption[]{The Feynman diagrams for the LNV processes $\mu^\pm \mu^\pm \rightarrow W_i^\pm W_j^\pm$ as those in the LRSM: figure (1, 2) are relating to the neutral Majorana lepton contributions and figure (3) is relating to the doubly charged Higgs contributions. Here the final sates $W_i^\pm W_j^\pm $ denote $W_1^\pm W_1^\pm $ or $W_1^\pm W_2^\pm $ or $W_2^\pm W_2^\pm $.}
\label{FllWW}
\end{figure}

The Feynman diagrams contributing to the LNV processes $\mu^\pm \mu^\pm \rightarrow W_i^\pm W_j^\pm, i,j=1,2$ at the tree-level are plotted in Fig.~\ref{FllWW}, where the final sates $W_i^\pm W_j^\pm $ denote $W_1^\pm W_1^\pm $ or $W_1^\pm W_2^\pm $ or $W_2^\pm W_2^\pm $. Summing up the fermions' spin and gauge bosons' polarizations, the squared amplitude for the processes $\mu^\pm \mu^\pm \rightarrow W_i^\pm W_j^\pm $ can be simplified by neglecting the terms proportional to $\mathcal{O}(\sin^2\zeta)$ and the charged lepton mass $m_\mu$ as
\begin{eqnarray}
&&|\mathcal{M}(\mu^\pm \mu^\pm \rightarrow W_1^\pm W_1^\pm )|^2\approx \frac{g_2^4}{M_{W_1}^4}\Big\{(|C_t^{11}|^2+|C_u^{11}|^2)M_{W_1}^2(2p_1\cdot k_1 p_1\cdot k_2+M_{W_1}^2 p_1\cdot p_2/2)\nonumber\\
&&\qquad\quad +2|C_t^{11}|^2k_1\cdot k_2(p_1\cdot k_1)^2+2|C_u^{11}|^2k_1\cdot k_2(p_1\cdot k_2)^2+\Re(C_t^{11}C_u^{11*})[2p_1\cdot p_2(k_1\cdot k_2)^2\nonumber\\
&&\quad\qquad +M_{W_1}^2(3M_{W_1}^2p_1\cdot p_2-4p_1\cdot k_1 p_1\cdot k_2)-2k_1\cdot k_2((p_1\cdot k_1)^2+(p_1\cdot k_2)^2)]\Big\},\\
&&|\mathcal{M}(\mu^\pm \mu^\pm \rightarrow W_1^\pm W_2^\pm )|^2\approx \frac{g_2^4}{2M_{W_1}^2M_{W_2}^2}\Big\{|C_t^{12}|^2[4M_{W_1}^2M_{W_2}^2 p_1\cdot k_1(p_2\cdot k_1-p_1\cdot p_2)\nonumber\\
&&\qquad\quad +8M_{W_1}^2 p_1\cdot k_1 p_2\cdot k_2(k_1\cdot k_2-p_1\cdot k_2)-M_{W_1}^4(M_{W_2}^2p_1\cdot p_2+2p_2\cdot k_2p_1\cdot k_2)\nonumber\\
&&\qquad\quad +4(p_1\cdot k_1)^2(M_{W_2}^2 p_1\cdot p_2+2p_2\cdot k_2 p_1\cdot k_2)]+|C_u^{12}|^2[4M_{W_1}^2M_{W_2}^2 p_1\cdot k_2(p_2\cdot k_2\nonumber\\
&&\qquad\quad -p_1\cdot p_2)+8M_{W_2}^2 p_1\cdot k_2 p_2\cdot k_1(k_1\cdot k_2-p_1\cdot k_1)-M_{W_2}^4(M_{W_1}^2p_1\cdot p_2\nonumber\\
&&\qquad\quad +2p_2\cdot k_1p_1\cdot k_1)+4(p_1\cdot k_2)^2(M_{W_1}^2 p_1\cdot p_2+2p_2\cdot k_1 p_1\cdot k_1)]\Big\},\\
&&|\mathcal{M}(\mu^\pm \mu^\pm \rightarrow W_2^\pm W_2^\pm )|^2\approx \frac{g_2^4}{M_{W_2}^4}\Big\{(|C_t^{22}|^2+|C_u^{22}|^2)M_{W_2}^2(2p_1\cdot k_1 p_1\cdot k_2+M_{W_2}^2 p_1\cdot p_2/2)\nonumber\\
&&\qquad\quad +2|C_t^{22}|^2k_1\cdot k_2(p_1\cdot k_1)^2+2|C_u^{22}|^2k_1\cdot k_2(p_1\cdot k_2)^2+\Re(C_t^{22}C_u^{22*})[2p_1\cdot p_2(k_1\cdot k_2)^2\nonumber\\
&&\quad\qquad +M_{W_2}^2(3M_{W_2}^2p_1\cdot p_2-4p_1\cdot k_1 p_1\cdot k_2)-2k_1\cdot k_2((p_1\cdot k_1)^2+(p_1\cdot k_2)^2)]\Big\},
\end{eqnarray}
where
\begin{eqnarray}
&&C_t^{11}=\cos^2\zeta(S_{2j})^2\frac{M_{N_j}}{t-M_{N_j}^2}+\frac{2\sqrt2Y_{L,22} v_L}{(s-M_{\Delta_L^{\pm\pm}}^2+
i\Gamma_{\Delta_L^{\pm\pm}}M_{\Delta_L^{\pm\pm}})},\nonumber\\
&&C_u^{11}=\cos^2\zeta(S_{2j})^2\frac{M_{N_j}}{u-M_{N_j}^2}+\frac{2\sqrt2Y_{L,22} v_L}{(s-M_{\Delta_L^{\pm\pm}}^2+
i\Gamma_{\Delta_L^{\pm\pm}}M_{\Delta_L^{\pm\pm}})},\label{31}\\
&&C_t^{12}=\cos^2\zeta (\frac{T_{2j}^* U_{2j}}{t-m_{\nu_j}^2}+\frac{V_{2j}^* S_{2j}}{t-M_{N_j}^2}),\nonumber\\
&&C_u^{12}=\cos^2\zeta (\frac{T_{2j}^* U_{2j}}{u-m_{\nu_j}^2}+\frac{V_{2j}^* S_{2j}}{t-M_{N_j}^2}),\label{32}\\
&&C_t^{22}=\cos^2\zeta(V_{2j}^*)^2\frac{M_{N_j}}{t-M_{N_j}^2}+\frac{2\sqrt2 Y_{R,22} v_R}{(s-M_{\Delta_R^{\pm\pm}}^2+
i\Gamma_{\Delta_R^{\pm\pm}}M_{\Delta_R^{\pm\pm}})},\nonumber\\
&&C_u^{22}=\cos^2\zeta(V_{2j}^*)^2\frac{M_{N_j}}{u-M_{N_j}^2}+\frac{2\sqrt2 Y_{R,22} v_R}{(s-M_{\Delta_R^{\pm\pm}}^2+
i\Gamma_{\Delta_R^{\pm\pm}}M_{\Delta_R^{\pm\pm}})}.\label{33}
\end{eqnarray}
Note that the mixing parameter $\sin \zeta$ is not present in Eqs.~(\ref{31},~\ref{32},~\ref{33}), because here in the calculations the further approximation, keeping the contributions up-to $\mathcal{O}(\sin\zeta)$ and setting the mass of the initial charged lepton to be zero is made. This approximation also disregards the contributions of doubly charged Higgs $\Delta_{L,R}^{\pm\pm}$ to the process $\mu^\pm\mu^\pm\rightarrow W_1^\pm W_2^\pm$. For {\bf TIII-NP}, Eq.~(\ref{31}) and Eq.~(\ref{33}) show that $\Delta_L^{\pm\pm}$ primarily contributes to the process $\mu^\pm \mu^\pm \to W_1^\pm W_1^\pm$, while $\Delta_R^{\pm\pm}$ primarily contributes to the process $\mu^\pm \mu^\pm \to W_2^\pm W_2^\pm$. Consequently, the distinct characteristics of $\Delta_L^{\pm\pm}$ and $\Delta_R^{\pm\pm}$ can be identified by observing the LNV di-boson processes at $\mu^\pm\mu^\pm$ colliders.

The results of $\mu^\pm\mu^\pm\to W_L^\pm W_L^\pm$ for {\bf TI-NP} can be acquired by setting $Y_{L,22}=0$ of Eq.~(\ref{31}), and the results of $\mu^\pm\mu^\pm\to W_1^\pm W_1^\pm$, $\mu^\pm\mu^\pm\to W_1^\pm W_2^\pm$, $\mu^\pm\mu^\pm\to W_2^\pm W_2^\pm$ for {\bf TII-NP} can be acquired by setting $Y_{L,22}=Y_{R,22}=0$ in Eqs.~(\ref{31},~\ref{33}). Based on the computations of the LNV di-boson processes, one may realize that the results of $\mu^\pm\mu^\pm\to W_1^\pm W_1^\pm$ cross section for {\bf TII-NP} is similar to the ones of $\mu^\pm\mu^\pm\to W_L^\pm W_L^\pm$ cross section for {\bf TI-NP}, the results of $\mu^\pm\mu^\pm\rightarrow W_1^\pm W_2^\pm$ cross section for {\bf TIII-NP} is similar to the ones of $\mu^\pm\mu^\pm\rightarrow W_1^\pm W_2^\pm$ cross section for {\bf TII-NP}. The cross sections of LNV di-boson processes can be written as
\begin{eqnarray}
&&\sigma=\frac{[(s-M_{W_i}^2-M_{W_j}^2)^2-4M_{W_i}^2M_{W_j}^2]^{1/2}}{32\pi s^2 A}\int_{-1}^1\frac{1}{4}|\mathcal{M}(\mu^\pm\mu^\pm\rightarrow W_i^\pm W_j^\pm )|^2 {\rm d}\cos\theta,\label{eq19}
\end{eqnarray}
where the factor $\frac{1}{4}$ comes from averaging the lepton spins in the initial state, $\theta$ is the angle between the momentum of outgoing $W_i$ and the collision axis. Considering the phase space integration factor of identified particles, $A=2$ for $W_iW_j=W_1W_1,\;W_2W_2$ and $A=1$ for $W_iW_j=W_1W_2$.

\section{Numerical results\label{sec4}}

In this section, we will calculate numerical results for the cross-sections of the LFV and LNV processes associated with {\bf TI-NP, TII-NP}, and {\bf TIII-NP}, using the formulas derived in Sec. III. To carry out the numerical evaluations, a lot of parameters in the NP which are constrained by available experiments need to be fixed, so let us explain how the parameters are fixed. The PDG~\cite{PDG} have collected a lot of the parameters such as $M_{W_L}(M_{W_1})=80.385\;{\rm GeV}$, $\Gamma_{W_L}=2.08\;{\rm GeV}$ and $\alpha_{em}(m_Z)=1/128.9$; the charged lepton masses $m_e=0.511\;{\rm MeV},\;m_\mu=0.105\;{\rm GeV},\;m_\tau=1.77\;{\rm GeV}$ etc, and we adopt them all. The sum of neutrino masses is limited in the range $\sum_i m_{\nu i}<0.12\;{\rm eV}$ by Plank~\cite{Esteban:2018azc}, it indicates that the contributions of the neutrino mass terms are negligible, regardless whether the neutrino masses are normal hierarchy ($m_{\nu_1}<m_{\nu_2}<m_{\nu_3}$) or inverse hierarchy ($m_{\nu_3}<m_{\nu_1}<m_{\nu_2}$). Hence we will ignore the contributions proportional to neutrino masses in the numerical evaluations. The matrix $U$ (the upper-left sub-matrix of the whole matrix $U_\nu$ in Eq.~(\ref{eqa3})) is taken as the PMNS mixing matrix~\cite{PDG} to describe the mixing of the light neutrinos.

For the light-heavy neutral lepton mixing (LHM) matrix $S_{ij}$, we set $S_{i}^2\equiv\sum_{j}|S_{ij}|^2\;(i,j=e,\mu,\tau)$ to describe the strength of LHM. The direct constraint on $S_{e}^2$ comes from the $0\nu2\beta$ decay searches as that $S_e^2\lesssim 10^{-5}$~\cite{Yang:2021ueh}. The direct constraint on $S_\mu^2$ is given by CMS as $S_\mu^2\lesssim 0.4 $ for TeV-scale heavy neutral lepton masses~\cite{CMS:2018iaf,ATLAS:2019kpx,CMS:2018jxx}. Recently, the constraint on $S_\mu^2$ from the future high-luminosity Large Hadron Collider (HL-LHC) is analyzed in Ref.~\cite{Fuks:2020att}, and the authors claim $S_\mu^2\lesssim 0.06$ for TeV-scale heavy neutral lepton masses. There is no direct constraint presently on $S_\tau^2$ for TeV-scale heavy neutral lepton masses. In the following analysis, we will show the proposed $\mu^\pm\mu^\pm$ colliders are more sensitive to $S_\mu^2$, $S_\tau^2$ than LHC and future HL-LHC for TeV-scale heavy neutral lepton masses, hence we set
\begin{eqnarray}
&&S_e^2\leq 10^{-5},\;S_\mu^2\leq0.01,\;S_\tau^2\leq0.01.\label{eq20}
\end{eqnarray}
where $S_\mu^2$ and $S_\tau^2$ are set a smaller value than that HL-LHC can reach at.

Considering the direct observations on the right-handed weak gauge boson $W_R$ ($W_2$), the lower bound for the mass of $W_2$ boson reads $M_{W_2}\gtrsim4.8\;{\rm TeV}$~\cite{Sirunyan:2017yrk,Sirunyan:2018pom,Aaboud:2018spl,Aaboud:2019wfg}, and its total decay width can be estimated as $\Gamma_{W_2}\approx0.028 M_{W_2}$~\cite{Mitra:2016kov} roughly. On the doubly charged Higgs masses, the most fresh limits from the LHC~\cite{ATLAS:2017iqw,CMS:2017pet} are $M_{\Delta_{L}^{\pm\pm}}\gtrsim800\;{\rm GeV},\;M_{\Delta_{R}^{\pm\pm}}\gtrsim650\;{\rm GeV}$. As pointed out above, the signals of $\Delta_L^{\pm\pm}$ and $\Delta_R^{\pm\pm}$ produced in the LFV di-lepton and LNV di-boson processes at $\mu^\pm\mu^\pm$ colliders can be used to identify them, and the signatures may well show their mass and width accordingly. According to Ref.~\cite{Keung:1984hn,Cahn:1988ru,Djouadi:1995gv,Aoki:2011pz}, the total decay widthes of $\Delta_L^{\pm\pm}$ and $\Delta_R^{\pm\pm}$ can be written as
\begin{eqnarray}
&&\Gamma_{\Delta_{L}^{--}}\approx \Gamma(\Delta_{L}^{--}\rightarrow l^-l^-)+\Gamma(\Delta_{L}^{--}\rightarrow W_1^-W_1^-)+...\nonumber\\
&&\qquad\;\approx\sum_{i=1}^3\frac{Y_{L,ii}^2 M_{\Delta_{L}^{--}}}{8\pi}+\frac{g_2^4 v_L^2}{16\pi M_{\Delta_L^{--}}}(3-\frac{M_{\Delta_L^{--}}^2}{M_{W_L}^2}+\frac{M_{\Delta_L^{--}}^4}{4M_{W_L}^4})\times
\sqrt{1-4\frac{M_{W_L}^2}{M_{\Delta_L^{--}}^2}}+...,\nonumber\\
&&\Gamma_{\Delta_{R}^{--}}\approx\Gamma(\Delta_{R}^{--}\rightarrow l^-l^-)+\Gamma(\Delta_{R}^{--}\rightarrow W_2^-W_2^{-(*)})+\Gamma(\Delta_{R}^{--}\rightarrow W_2^-W_2^-)+...\nonumber\\
&&\qquad\;\approx \sum_{i=1}^3\frac{Y_{R,ii}^2 M_{\Delta_{R}^{--}}}{8\pi}+\Gamma(\Delta_{R}^{--}\rightarrow W_2^-W_2^{-(*)})+\Gamma(\Delta_{R}^{--}\rightarrow W_2^-W_2^-)+...,\label{decay0}
\end{eqnarray}
where
\begin{eqnarray}
&&\Gamma(\Delta_{R}^{--}\rightarrow W_2^-W_2^{-(*)})\approx\frac{g_2^6 v_R^2M_{\Delta_R^{--}}}{128\pi^3 M_{W_R}^2}F(\frac{M_{W_R}^2}{M_{\Delta_R^{--}}^2}),\nonumber\\
&&\Gamma(\Delta_{R}^{--}\rightarrow W_2^-W_2^-)\approx\frac{g_2^4 v_R^2}{16\pi M_{\Delta_R^{--}}}(3-\frac{M_{\Delta_R^{--}}^2}{M_{W_R}^2}+\frac{M_{\Delta_R^{--}}^4}{4M_{W_R}^4})\times
\sqrt{1-4\frac{M_{W_R}^2}{M_{\Delta_R^{--}}^2}},\label{decay1}
\end{eqnarray}
and
\begin{eqnarray}
&&F(x)=-|1-x|(\frac{47}{2}x-\frac{13}{2}+\frac{1}{x})+3(1-6x+4x^2)|\log\sqrt x|+\nonumber\\
&&\qquad\quad\frac{3(1-8x+20x^2)}{\sqrt{4x-1}}\arccos(\frac{3x-1}{2x^{3/2}}).\label{decay2}
\end{eqnarray}
The relevant Yukawa coupling $Y_{R,ll}$ is not a free parameter\footnote{When $M_{W_2}$, $M_{N_l} (l=1,2,3)$ etc are fixed, then according to Eqs.~(\ref{eq4},~\ref{eq7}-\ref{eq9}) the Yukawa coupling $Y_R$ is not a free parameter, namely in numerical calculations we will use the other relevant parameters to replace $Y_R$.}, therefore we only have the Yukawa coupling $Y_L$ of the left-handed doubly charged Higgs to the leptons need to be set. Generally it take the formulation below:
\begin{eqnarray}
Y_L={\rm diag}\;(Y_{ee},\;Y_{\mu\mu},\;Y_{\tau\tau}). \label{Y-matrix}
\end{eqnarray}
$Y_{ee}$ is constrained strongly by the $0\nu2\beta$ decay experiments in the range $Y_{ee}\lesssim0.04$. In addition, a small VEV $v_L$ of $\Delta_L^0$, $v_L\lesssim 5.0\;{\rm GeV}$, is constrained by the $\rho-$parameter~\cite{PDG}, so later on we will set it as $v_L=0.1\;{\rm GeV}$ to simplify the numerical evaluations.

\subsection{Numerical results for the LFV di-lepton processes}

In this subsection, the numerical results about the LFV di-lepton processes $\mu^\pm\mu^\pm\to e^\pm e^\pm$ and $\mu^\pm\mu^\pm\to \tau^\pm \tau^\pm$ for {\bf TI-NP}, {\bf TII-NP}, {\bf TIII-NP} are presented. The characters of LFV di-lepton processes are clear and practically free from SM background at high energy same-sign muon colliders~\cite{Cannoni:2002ny}\footnote{The LFV di-lepton processes, being two-two body processes, are very different from the SM processes. In SM the processes with two same flavor charged leptons in initiate state and two same but different from initial state flavor charged leptons in final state must be of mixing neutrino and, that the behavior is very different from those concerned here.}. Firstly, let us focus lights on the effects of LHM parameter $S_\mu^2$ and the heavy neutral lepton mass $M_{N_2}$.

\begin{figure}
\setlength{\unitlength}{1mm}
\centering
\includegraphics[width=2.7in]{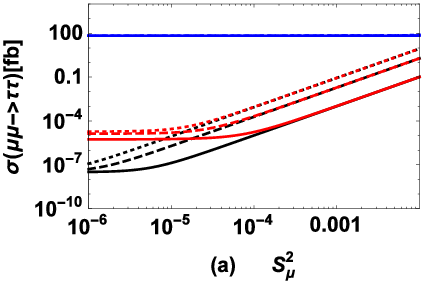}
\vspace{0.5cm}
\includegraphics[width=2.7in]{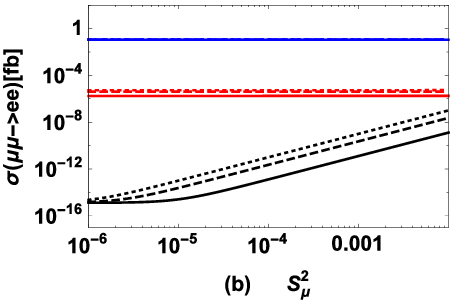}
\vspace{0cm}
\caption[]{The cross-section $\sigma(\mu^\pm\mu^\pm\rightarrow l^\pm l^\pm)$ versus $S_\mu^2$ for $M_{N_1}=1.0\;{\rm TeV}$, $M_{N_3}=3.0\;{\rm TeV}$, $S_e^2=10^{-5},\;S_\tau^2=0.01$, $\sqrt s=5.0\;{\rm TeV}$, where figure (a) for $l=\tau$ and figure (b) for $l=e$. The solid, dashed, dotted curves denote the results for $M_{N_2}=1.0,\;2.0,\;3.0\;$TeV respectively, the black curves denote the results for {\bf TI-NP}, the red curves denote the results for {\bf TII-NP} with $M_{W_2}=5.0\;{\rm TeV}$, the blue curves denote the results for {\bf TIII-NP} with $M_{W_2}=5.0\;{\rm TeV}$, $M_{\Delta_L^{\pm\pm}}=10.0\;$TeV, $M_{\Delta_R^{\pm\pm}}=11.0\;$TeV, $Y_{ee}=0.04$, $Y_{\mu\mu}=1.0$, $Y_{\tau\tau}=1.0$.}
\label{LFV1}
\end{figure}

\begin{figure}
\setlength{\unitlength}{1mm}
\centering
\includegraphics[width=2.7in]{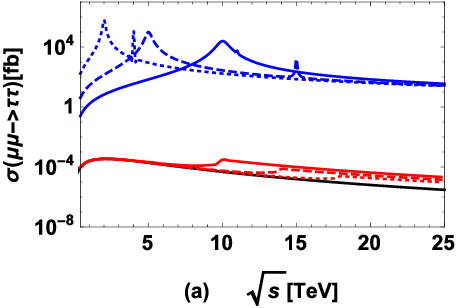}
\vspace{0.5cm}
\includegraphics[width=2.7in]{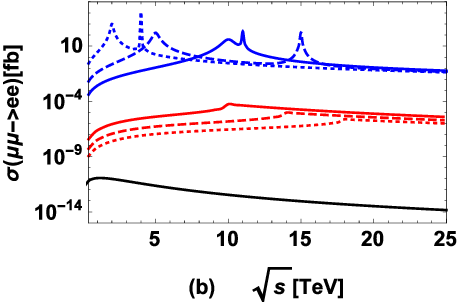}
\vspace{0cm}
\caption[]{$\sigma(\mu^\pm\mu^\pm\to l^\pm l^\pm)$ versus $\sqrt s$ with $M_{N_1}=1.0\;{\rm TeV}$, $M_{N_2}=2.0\;{\rm TeV}$, $M_{N_3}=3.0\;{\rm TeV}$, $S_e^2=10^{-5}$, $S_\mu^2=10^{-4}$, $S_\tau^2=10^{-2}$, where figure (a) for $l=\tau$ and Figure (b) for $l=e$. The black curves denote the results for {\bf TI-NP}. The red curves denote the results for {\bf TII-NP}, where the red solid, red dashed, red dotted curves denote the results for $M_{W_2}=5.0,\;7.0,\;9.0\;$TeV respectively. The blue curves denote the results for {\bf TIII-NP} with $M_{W_2}=5.0\;$TeV, $Y_{ee}=0.04$, $Y_{\mu\mu}=1.0$, $Y_{\tau\tau}=1.0$, and the blue solid curves denote the results for $M_{\Delta_L^{\pm\pm}}=10.0\;$TeV, $M_{\Delta_R^{\pm\pm}}=11.0\;$TeV, the blue dashed curves denote the results for $M_{\Delta_L^{\pm\pm}}=5.0\;$TeV, $M_{\Delta_R^{\pm\pm}}=15.0\;$TeV, the blue dotted curves denote the results for setting $M_{\Delta_L^{\pm\pm}}=2.0\;$TeV, $M_{\Delta_R^{\pm\pm}}=4.0\;$TeV.}
\label{LFV2}
\end{figure}

Taking possible $M_{N_1}=1.0\;{\rm TeV}$, $M_{N_3}=3.0\;{\rm TeV}$, $S_e^2=10^{-5},\;S_\tau^2=10^{-2}$, $\sqrt s=5.0\;{\rm TeV}$ as an example, we present the results of  $\sigma(\mu^\pm\mu^\pm\rightarrow \tau^\pm\tau^\pm)$, $\sigma(\mu^\pm\mu^\pm\rightarrow e^\pm e^\pm)$ versus $S_\mu^2$ in Fig.~\ref{LFV1} (a) and Fig.~\ref{LFV1} (b) accordingly, where the solid, dashed, dotted curves denote the results for $M_{N_2}=1.0,\;2.0,\;3.0\;$TeV respectively. In the figures the black curves denote the results for {\bf TI-NP}, the red curves denote the results for {\bf TII-NP} with $M_{W_2}=5.0\;{\rm TeV}$, the blue curves denote the results for {\bf TIII-NP} with $M_{W_2}=5.0\;{\rm TeV}$, $M_{\Delta_L^{\pm\pm}}=10.0\;$TeV, $M_{\Delta_R^{\pm\pm}}=11.0\;$TeV, $Y_{ee}=0.04$, $Y_{\mu\mu}=1.0$, $Y_{\tau\tau}=1.0$. Note here that the masses of the heavy neutral leptons are set at the TeV scale, because it is hard to be excluded by experiments in near future. We additionally try to set $\sqrt s=5.0\;{\rm TeV}$ as a representative collision energy for TeV scale $\mu^\pm\mu^\pm$ colliders in Fig.~\ref{LFV1}, and the numerical results for various $\sqrt s$ are investigated in Fig.~\ref{LFV2}.

The numerical results in Fig.~\ref{LFV1} indicate that when an integrated luminosity
of $500\;{\rm fb}^{-1}$ is accumulated at a TeV scale $\mu^\pm \mu^\pm $ collider, the process $\mu^\pm\mu^\pm\rightarrow \tau^\pm
\tau^\pm$ predicted by {\bf TI-NP}, {\bf TII-NP} and the processes $\mu^\pm\mu^\pm\rightarrow \tau^\pm \tau^\pm$,
$\mu^\pm\mu^\pm\rightarrow e^\pm e^\pm$ predicted by {\bf TIII-NP} have great opportunities to be observed. Because their cross
section can be larger than $0.1\;{\rm fb}$ in a reasonable chosen parameter space, that means more than $50$ signal events/year
can be collected, so it indicates that the $\mu^\pm\mu^\pm$ collider is more sensitive to $S_\mu^2$ and $S_\tau^2$ than the future HL-LHC. The contributions from Majorana neutral leptons to $\sigma(\mu^\pm\mu^\pm\rightarrow \tau^\pm \tau^\pm)$, $\sigma(\mu^\pm\mu^\pm\rightarrow e^\pm e^\pm)$ are proportional to $S_\mu^2 S_\tau^2$, $S_\mu^2 S_e^2$ respectively. And the contributions from doubly charged Higgs to $\sigma(\mu^\pm\mu^\pm\rightarrow \tau^\pm \tau^\pm)$, $\sigma(\mu^\pm\mu^\pm\rightarrow e^\pm e^\pm)$ are proportional to $Y_{\mu\mu}Y_{\tau\tau}$, $Y_{\mu\mu}Y_{ee}$ respectively. Moreover, $S_e^2$ and $Y_{ee}$ are limited to be small by the $0\nu2\beta$ experiments while $S_\tau^2$ and $Y_{\tau\tau}$ are not, which leads to the contributions to $\sigma(\mu^\pm\mu^\pm\rightarrow e^\pm e^\pm)$ are suppressed. The fact can be see clearly by comparing Fig.~\ref{LFV1} (a) with Fig.~\ref{LFV1} (b). From the figures one may see that the three blue curves merge together in the logarithmic coordinate, and the predicted cross-sections $\sigma(\mu^\pm \mu^\pm \rightarrow \tau^\pm \tau^\pm )$, $\sigma(\mu^\pm \mu^\pm \to e^\pm e^\pm )$ for {\bf TIII-NP} are much larger than the ones predicted by {\bf TI-NP} and {\bf TII-NP}. It is because that the leading contributions to the LFV di-lepton processes for {\bf TIII-NP} come from the $s-$channel mediation of the doubly charged Higgs at the tree level, while the ones for {\bf TI-NP} and {\bf TII-NP} start with the one-loop level which involves the Majorana neutral leptons. The black and red curves in Fig.~\ref{LFV1} (a), (b) show that $\sigma(\mu^\pm \mu^\pm \rightarrow \tau^\pm \tau^\pm )$ and $\sigma(\mu^\pm \mu^\pm \rightarrow e^\pm e^\pm )$ are dominated by right-handed gauge boson for {\bf TII-NP} if $S_l^2$ ($l=e,\;\mu,\;\tau$) are small. The results of $\sigma(\mu^\pm \mu^\pm \to \tau^\pm \tau^\pm )$ for {\bf TI-NP}, {\bf TII-NP} are similar when $S_\mu^2\gtrsim 10^{-4}$, and $\sigma(\mu^\pm \mu^\pm \to e^\pm e^\pm )$ for {\bf TII-NP} is always larger than that for {\bf TI-NP} owing to the fact that there are only the contributions from $W_L$ mediation for {\bf TI-NP}. The contributions to the LFV di-lepton processes for {\bf TIII-NP} are dominated by the doubly charged Higgs so the dependence on $S_l^2$ may be ignorable. A large $M_{N_2}$ plays an enhancing role on $\sigma(\mu^\pm \mu^\pm \to \tau^\pm \tau^\pm )$ and $\sigma(\mu^\pm \mu^\pm \to e^\pm e^\pm )$ for {\bf TI-NP} and {\bf TII-NP}, because the goldstone component (in Feynman gauge) makes dominant contributions to the LFV di-lepton processes for {\bf TI-NP} and {\bf TII-NP}, the corresponding couplings increase with the increasing of heavy neutral lepton masses for a given $S_l^2$.

The results of $\sigma(\mu^\pm\mu^\pm\to \tau^\pm\tau^\pm)$, $\sigma(\mu^\pm\mu^\pm\to e^\pm e^\pm)$ versus the $\sqrt s$ are presented in Fig.~\ref{LFV2} (a) and in Fig.~\ref{LFV2} (b) respectively for $M_{N_1}=1.0\;{\rm TeV}$, $M_{N_2}=2.0\;{\rm TeV}$, $M_{N_3}=3.0\;{\rm TeV}$, $S_e^2=10^{-5}$, $S_\mu^2=10^{-4}$, $S_\tau^2=10^{-2}$. The black curves denote the results for {\bf TI-NP}. The red curves denote the results for {\bf TII-NP}, where the red solid, red dashed, red dotted curves denote the results when $M_{W_2}=5.0,\;7.0,\;9.0\;$TeV respectively. The blue curves denote the results for {\bf TIII-NP} with possible parameters $M_{W_2}=5.0\;$TeV, $Y_{ee}=0.04$, $Y_{\mu\mu}=1.0$, $Y_{\tau\tau}=1.0$, where the blue solid curves denote the results with $M_{\Delta_L^{\pm\pm}}=10.0\;$TeV, $M_{\Delta_R^{\pm\pm}}=11.0\;$TeV, the blue dashed curves denote the results with $M_{\Delta_L^{\pm\pm}}=5.0\;$TeV, $M_{\Delta_R^{\pm\pm}}=15.0\;$TeV, the blue dotted curves denote the results with $M_{\Delta_L^{\pm\pm}}=2.0\;$TeV, $M_{\Delta_R^{\pm\pm}}=4.0\;$TeV. As for {\bf TIII-NP}, since the contributions to the LFV di-lepton processes are dominated by doubly charged Higgs, hence we do not present the results for various $M_{W_2}$.

\begin{figure}
\setlength{\unitlength}{1mm}
\centering
\includegraphics[width=2.5in]{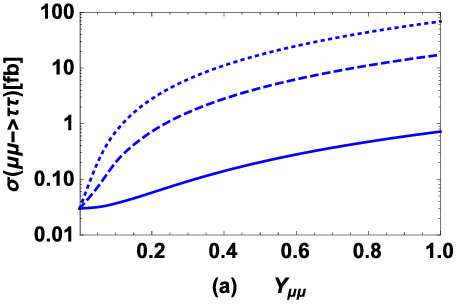}
\vspace{0.5cm}
\includegraphics[width=2.5in]{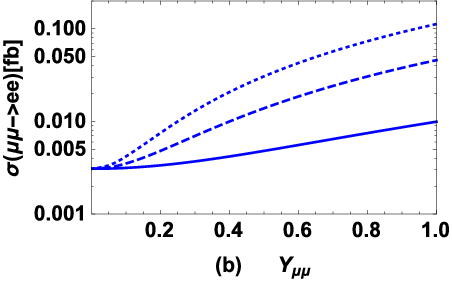}
\vspace{0cm}
\caption[]{The cross-section $\sigma(\mu^\pm \mu^\pm \rightarrow l^\pm l^\pm )$ versus $Y_{\mu\mu}$ for $\sqrt s=5.0\;{\rm TeV}$, $M_{W_2}=5.0\;{\rm TeV}$, $M_{N_1}=1.0\;{\rm TeV}$, $M_{N_2}=2.0\;{\rm TeV}$, $M_{N_3}=3.0\;{\rm TeV}$, $M_{\Delta_L^{\pm\pm}}=10.0\;{\rm TeV}$, $M_{\Delta_R^{\pm\pm}}=11.0\;{\rm TeV}$, $S_e^2=10^{-5}$, $S_\mu^2=10^{-4}$, $S_\tau^2=10^{-2}$. $l=\tau$ for figure (a), and the blue solid, blue dashed, blue dotted blue curves denote the results for $Y_{\tau\tau}=0.1,\;0.5,\;1.0$ respectively. $l=e$ for figure (b), and the blue solid, blue dashed, blue dotted curves denote the results for $Y_{ee}=0.01,\;0.025,\;0.04$ respectively.}
\label{LFV3}
\end{figure}

The small ``hill'' in red curves in Fig.~\ref{LFV2} is the result due to the $W_2$ boson being on-shell. Fig.~\ref{LFV2} (b) shows the fact that the cross-section $\sigma(\mu^\pm \mu^\pm \rightarrow e^\pm e^\pm )$ for {\bf TII-NP} is always larger than the one for {\bf TI-NP} as analyzed above. The explicit resonance enhancements (the peaks) appearing on the blue curves for {\bf TIII-NP} in Fig.~\ref{LFV2} are owing to $\sqrt s$ crossing the doubly charged Higgs $\Delta_L^{\pm\pm}$ or $\Delta_R^{\pm\pm}$ mass value as $\sqrt{s}$ increasing, it indicates the LFV processes $\mu^\pm\mu^\pm\rightarrow \tau^\pm \tau^\pm$ and $\mu^\pm\mu^\pm\to e^\pm e^\pm$ are very good channels to observe the two doubly charged Higgs at a $\mu^\pm \mu^\pm$ collider by scanning the collision energy. Namely if the resonance enhancements in the LFV processes appear, it means that the signals may be used to observe the doubly charged Higgs $\Delta_L^{\pm\pm}$ and $\Delta_R^{\pm\pm}$. The enhancement signal of the doubly charged Higgs depends on their total widthes, and the height of the resonance peak depends on the relevant Yukawa coupling $Y_{ll}$ (see Eq.~(\ref{Y-matrix})) of the doubly charged Higgs to the leptons. And how the coupling $Y_{ll}$ affects the cross sections $\sigma(\mu^\pm \mu^\pm \to \tau^\pm \tau^\pm )$ and $\sigma(\mu^\pm \mu^\pm \to e^\pm e^\pm )$ are computed and presented in Fig.~\ref{LFV3}.

In the plotting of Fig.~\ref{LFV3}, we set $\sqrt s=5.0\;{\rm TeV}$, $M_{W_2}=5.0\;{\rm TeV}$ (in fact for {\bf TIII-NP} $M_{W_2}$ affects the numerical results of LFV processes slightly, hence fixing the value of $M_{W_2}$ does not lose the general features which we are interested in), $M_{N_1}=1.0\;{\rm TeV}$, $M_{N_2}=2.0\;{\rm TeV}$, $M_{N_3}=3.0\;{\rm TeV}$, $M_{\Delta_L^{\pm\pm}}=10.0\;{\rm TeV}$, $M_{\Delta_R^{\pm\pm}}=11.0\;{\rm TeV}$, $S_e^2=10^{-5}$, $S_\mu^2=10^{-4}$, $S_\tau^2=10^{-2}$. Fig.~\ref{LFV3}~(a) is for $l=\tau$ and the blue solid, blue dashed, blue dotted curves denote the results with various $Y_{\tau\tau}=0.1,\;0.5,\;1.0$ respectively. Fig.~\ref{LFV3}~(b) is for $l=e$ and the blue solid, blue dashed, blue dotted curves denote the results with various $Y_{ee}=0.01,\;0.025,\;0.04$ respectively. The contributions from doubly charged Higgs are proportional to ($Y_{\mu\mu}\cdot Y_{\tau\tau}$) for $\mu^\pm \mu^\pm \rightarrow \tau^\pm \tau^\pm $ and ($Y_{\mu\mu}\cdot Y_{ee}$) for $\mu^\pm \mu^\pm \rightarrow e^\pm e^\pm $, which indicates that $\sigma(\mu^\pm \mu^\pm \rightarrow \tau^\pm \tau^\pm )$ increases with increasing of $Y_{\mu\mu}$ and $Y_{\tau\tau}$; the cross-section $\sigma(\mu^\pm \mu^\pm \rightarrow e^\pm e^\pm )$ increases with the increasing of $Y_{\mu\mu}$ and $Y_{ee}$. And the behavior shown in Fig.~\ref{LFV3} is that as expects indeed. Comparing the bound on relevant couplings from the muonium to anti-muonium transition experiment $(Y_{ee}Y_{\mu\mu})/(4\sqrt 2 M_{\Delta_L^{\pm\pm}}^2 G_F)\leq 3\times10^{-3}$~\cite{Willmann:1998gd}, if $\sigma(\mu^\pm \mu^\pm \rightarrow e^\pm e^\pm )\leq 1\;{\rm fb}$ can be set by the future $\mu^\pm \mu^\pm$ collider, the bound $(Y_{ee}Y_{\mu\mu})/(4\sqrt 2 M_{\Delta_L^{\pm\pm}}^2 G_F)\leq6.15\times10^{-6}$  will be set which enhances the bound obtained by the muonium to anti-muonium transition experiment about three orders of magnitude.

\begin{figure}
\setlength{\unitlength}{1mm}
\centering
\includegraphics[width=2.5in]{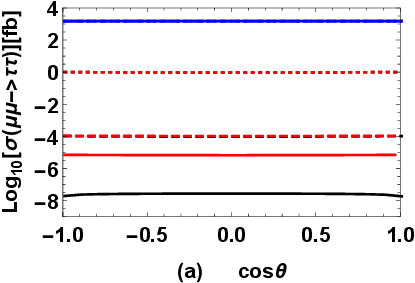}
\vspace{0.5cm}
\includegraphics[width=2.5in]{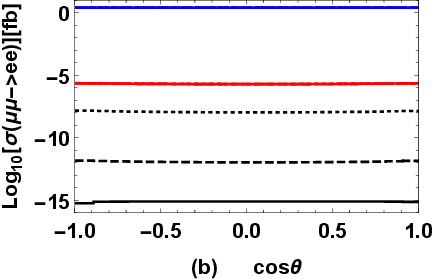}
\vspace{0cm}
\caption[]{The angle distributions of the process $\mu^\pm \mu^\pm \rightarrow \tau^\pm \tau^\pm $ for $M_{N_1}=1\;{\rm TeV}$, $M_{N_2}=2\;{\rm TeV}$, $M_{N_3}=3\;{\rm TeV}$, $S_e^2=10^{-5}$, $S_\tau^2=10^{-2}$, $\sqrt s=5\;{\rm TeV}$, where $l=\tau$ for (a), $l=e$ for (b), and the solid, dashed, dotted curves denote the results with $S_\mu^2=10^{-6},\;10^{-4},\;10^{-2}$ respectively. The black curves denote the results for {\bf TI-NP}, the red curves denote the results for {\bf TII-NP} with $M_{W_2}=5\;{\rm TeV}$, the blue curves denote results for {\bf TIII-NP} with $M_{W_2}=5\;{\rm TeV}$, $M_{\Delta^{\pm\pm}}=3\;$TeV, $Y_{ee}=0.04$, $Y_{\mu\mu}=1$, $Y_{\tau\tau}=1$.}
\label{LFV4}
\end{figure}
Finally, to show the characters of the LFV di-lepton processes, we further take the typical parameters $M_{N_1}=1\;{\rm TeV}$, $M_{N_2}=2\;{\rm TeV}$, $M_{N_3}=3\;{\rm TeV}$, $S_e^2=10^{-5}$, $S_\tau^2=10^{-2}$, $\sqrt s=5\;{\rm TeV}$, to calculate the angle distributions of the processes $\mu^\pm \mu^\pm \rightarrow \tau^\pm \tau^\pm $, $\mu^\pm \mu^\pm \rightarrow e^\pm e^\pm $ and plot the results in Fig.~\ref{LFV4} (a), Fig.~\ref{LFV4} (b) respectively. Here in the figures the solid, dashed, dotted curves denote the results with $S_\mu^2=10^{-6},\;10^{-4},\;10^{-2}$ respectively; the black curves denote the results for {\bf TI-NP}, the red curves denote the results for {\bf TII-NP} with $M_{W_2}=5\;{\rm TeV}$, the blue curves denote results for {\bf TIII-NP} with $M_{W_2}=5\;{\rm TeV}$, $M_{\Delta^{\pm\pm}}=3\;$TeV, $Y_{ee}=0.04$, $Y_{\mu\mu}=1$, $Y_{\tau\tau}=1$. The three blue curves merge together within the plotting scale of Fig.~\ref{LFV4} (a), Fig.~\ref{LFV4} (b). The black dashed, black dotted curves merge into the red dashed, red dotted curves respectively within the plotting scale of Fig.~\ref{LFV4} (a). The three red curves merge together at the plotting scale of Fig.~\ref{LFV4} (b). The picture shows that the angular distributions of the processes $\mu^\pm \mu^\pm \rightarrow \tau^\pm \tau^\pm $, $\mu^\pm \mu^\pm \rightarrow e^\pm e^\pm $ are flat in these three types of NP models.

\subsection{Numerical results for the LNV di-boson processes}

In this subsection, we compute the LNV di-boson processes $\mu^\pm\mu^\pm\to W^\pm_{i}W^\pm_{j}, (i,j=1,2)$ for {\bf TI-NP}, {\bf TII-NP}, {\bf TIII-NP} and present the numerical results properly. For the collider search of $\mu^-\mu^-\to W^-W^-$ (the case of $\mu^+\mu^+\to W^+W^+$ collisions is similar), the different decay channels of the final $W$ bosons corresponding to different SM background processes. For the pure leptonic channel $\mu^-\mu^-+E\!\!\!\!/_T$ and $\mu^-e^-+E\!\!\!\!/_T$ where $E\!\!\!\!/_T$ denotes the missing energy carried by neutrinos, we compute the SM background processes $\mu^-\mu^- \rightarrow W^- \mu^-\nu_\mu $ and $\mu^-\mu^- \rightarrow Z \mu^-\mu^- $ in terms of MadGraph5~\cite{Alwall:2014hca}, and the results at $\sqrt s=15\;{\rm TeV}$ are so large as
\begin{eqnarray}
&&\sigma(\mu^- \mu^- \rightarrow W^- \mu^-\nu_\mu)\approx325.6\pm0.9\;{\rm fb},\;\sigma(\mu^- \mu^- \rightarrow Z \mu^-\mu^-)\approx 5.797\pm0.019\;{\rm fb}.
\end{eqnarray}
It indicates they may make significant backgrounds when the pure leptonic channel $\mu^-\mu^-+E\!\!\!\!/_T$, $\mu^-e^-+E\!\!\!\!/_T$ are considered in observing the LNV processes, and the signals for the LNV processes are very hard to be picked up from the backgrounds~\cite{Wang:2016eln}. However there is a technique way to avoid the problem at least, namely ignoring all the events in which the decays $W^-\to \mu^- \bar{\nu}_\mu$ and/or $W^-\to \tau^- \bar{\nu}_\tau$ with $\tau^- \to \mu^-\nu_\tau\bar{\nu}_\mu$ are involved, and taking into account only the events in which the $W^-$ bosons decay either to $e \bar{\nu}_e$ or $\tau \bar\nu_\tau$ (except the $\tau$ lepton decay $\tau^-\to \mu^-\nu_\tau \bar{\nu}_\mu$) or quarks ($q\bar{q'}$). In this way, although the efficiency of identifying $W$-boson(s) in the final states of the LNV processes will be lost a bit, the accuracy for identifying the LNV processes will be ensured. Therefore later on we will not concern this kind of possible SM backgrounds for the LNV any more. As analyzed in Ref.~\cite{Wang:2016eln} (which focus on the same-sign electron colliders and the case of same-sign muon colliders considered in this work is similar), the dominant SM background process is $\mu^-\mu^-\rightarrow W^-W^-\nu_\mu\nu_\mu$ for the pure leptonic channel $e^-e^-+E\!\!\!\!/_T$. The $W$ bosons in the final state have effective missing energies and momenta due to the involving neutrinos in the final state, it indicates that the background process can be highly suppressed by cutting the invariant-mass of the two outgoing electrons. In addition, it is concluded in Ref.~\cite{Wang:2016eln} that the semi-leptonic channel $e^-+j_W+E\!\!\!\!/_T$ and the pure hadronic channel $2j_W$ also have great potential to observe the signal processes.

We stated in Sec.~\ref{sec3} that the cross section of $\mu^\pm\mu^\pm\rightarrow W_1^\pm W_1^\pm$ for {\bf TII-NP} is similar to the one of $\mu^\pm\mu^\pm\rightarrow W_L^\pm W_L^\pm$ for {\bf TI-NP}, and the one of $\mu^\pm\mu^\pm\rightarrow W_1^\pm W_2^\pm$ for {\bf TIII-NP} is similar to the one of $\mu^\pm\mu^\pm\rightarrow W_1^\pm W_2^\pm$ for {\bf TII-NP}, therefore for simplicity and avoiding repeats, below we will not present the results about $\mu^\pm\mu^\pm\rightarrow W_1^\pm W_1^\pm$ for {\bf TII-NP} and the results about $\mu^\pm\mu^\pm\rightarrow W_1^\pm W_2^\pm$ for {\bf TIII-NP}.

\begin{figure}
\setlength{\unitlength}{1mm}
\centering
\includegraphics[width=2.5in]{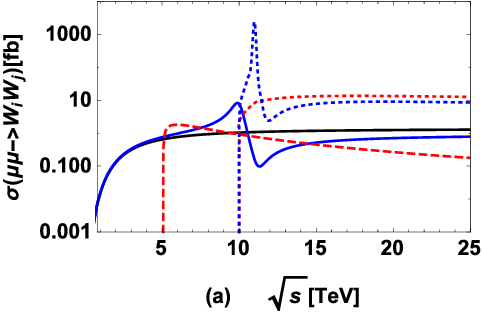}
\vspace{0.5cm}
\includegraphics[width=2.5in]{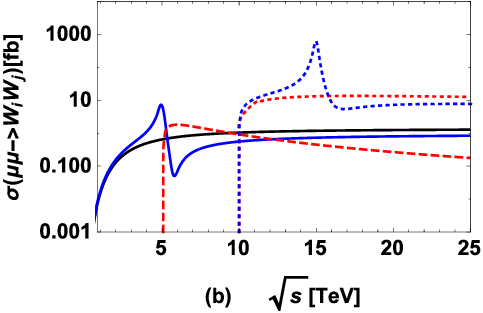}
\vspace{0cm}
\caption[]{$\sigma(\mu^\pm \mu^\pm \rightarrow W_i^\pm W_j^\pm)$ versus $\sqrt s$ with $M_{N_2}=2.0\;{\rm TeV}$, $S_\mu^2=10^{-4}$, where the solid, dashed, dotted curves denote the results for the processes when $W_iW_j=W_1W_1$, $W_1W_2$, $W_2W_2$ respectively. Figure (a) is about those additionally to have $M_{\Delta_L^{\pm\pm}}=10.0\;{\rm TeV}$, $M_{\Delta_R^{\pm\pm}}=11.0\;{\rm TeV}$ for {\bf TIII-NP}. Figure (b) is about those additionally to have $M_{\Delta_L^{\pm\pm}}=5.0\;{\rm TeV}$, $M_{\Delta_R^{\pm\pm}}=15.0\;{\rm TeV}$ for  {\bf TIII-NP}. The black curves denote the results for {\bf TI-NP}, the red curves denote the results for {\bf TII-NP} with $M_{W_2}=5.0\;$TeV, the blue curves denote the results for {\bf TIII-NP} with $M_{W_2}=5.0\;$TeV, $Y_{\mu\mu}=1.0$.}
\label{LNV1}
\end{figure}

The results on the cross-sections $\sigma(\mu^\pm \mu^\pm \rightarrow W_i^\pm W_j^\pm),\, (i,j=1,2)$ versus $\sqrt s$ are presented in Fig.~\ref{LNV1} for $M_{N_2}=2.0\;{\rm TeV}$, $S_\mu^2=10^{-4}$, where the solid, dashed, dotted curves denote the results for the processes when $W_iW_j=W_1W_1$, $W_1W_2$, $W_2W_2$ respectively. Fig.~\ref{LNV1} (a) is about those, additionally to have $M_{\Delta_L^{\pm\pm}}=10.0\;{\rm TeV}$, $M_{\Delta_R^{\pm\pm}}=11.0\;{\rm TeV}$ for {\bf TIII-NP}. Fig.~\ref{LNV1} (b) is about those, additionally to have $M_{\Delta_L^{\pm\pm}}=5.0\;{\rm TeV}$, $M_{\Delta_R^{\pm\pm}}=15.0\;{\rm TeV}$ for {\bf TIII-NP}. The black curve denotes the results for {\bf TI-NP}, the red curves denote the results for {\bf TII-NP} with $M_{W_2}=5.0\;$TeV, the blue curves denote the results for {\bf TIII-NP} with $M_{W_2}=5.0\;$TeV, $Y_{\mu\mu}=1.0$.

In Fig.~\ref{LNV1} the blue dotted curve and the red dotted curve merge together at $\sqrt s \simeq 10.0\;$TeV, because the process $\mu^\pm \mu^\pm \rightarrow W_2^\pm W_2^\pm$ starts from $\sqrt s=10.0\;$TeV when $M_{W_2}=5.0\;$TeV. The enhancement as a hill on blue curve for $\sigma(\mu^\pm \mu^\pm \rightarrow W_1^\pm W_1^\pm )$ or $\sigma(\mu^\pm \mu^\pm \rightarrow W_2^\pm W_2^\pm )$ is the resonance signal due to the $s-$channel $\Delta_L^{\pm\pm}$ or $\Delta_R^{\pm\pm}$, and it corresponds explicitly to the blue solid and blue dotted curves respectively. From the figure one may see the resonance signal which is caused by $\Delta_L^{\pm\pm}$ or $\Delta_R^{\pm\pm}$ i.e. the resonance peak appears in the process $\mu^\pm \mu^\pm \rightarrow W_1^\pm W_1^\pm$ or in the process $\mu^\pm \mu^\pm \rightarrow W_2^\pm W_2^\pm$ at $\mu^\pm\mu^\pm$ colliders. The observed resonance enhancement appears in $\mu^\pm \mu^\pm \rightarrow W_1^\pm W_1^\pm$ when the $s-$channel $\Delta_L^{\pm\pm}$ plays roles, and the observed resonance enhancement appears in $\mu^\pm \mu^\pm \rightarrow W_2^\pm W_2^\pm$ when the $s-$channel $\Delta_R^{\pm\pm}$ plays roles. In addition, in Fig.~\ref{LNV1} the `valley' which appears on the blue solid or blue dotted curve due to the interference effect between the contributions from Majorana neutral leptons and those from the doubly charged Higgs.

\begin{figure}
\setlength{\unitlength}{1mm}
\centering
\includegraphics[width=2.5in]{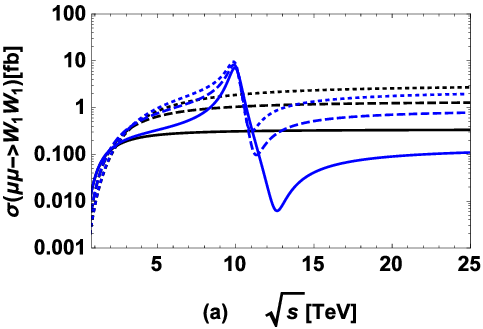}
\vspace{0.5cm}
\includegraphics[width=2.5in]{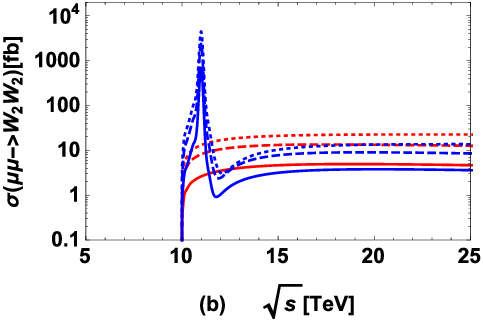}
\vspace{0cm}
\caption[]{$\sigma(\mu^\pm \mu^\pm \rightarrow W_i^\pm W_j^\pm)$ versus $\sqrt s$ with $S_\mu^2=10^{-4}$, where figure (a) is for $W_iW_j=W_1W_1$, figure (b) is for $W_iW_j=W_2W_2$. The solid, dashed, dotted curves in the figures denote the results for $M_{N_2}=1.0,\;2.0,\;3.0\;{\rm TeV}$ respectively. The black curves denote the results for {\bf TI-NP}, the red curves denote the results for {\bf TII-NP} with $M_{W_2}=5.0\;{\rm TeV}$. The blue curves denote the results for {\bf TIII-NP} with $M_{W_2}=5.0\;{\rm TeV}$, $Y_{\mu\mu}=1.0$, $M_{\Delta_L^{\pm\pm}}=10.0\;{\rm TeV}$, $M_{\Delta_R^{\pm\pm}}=11.0\;{\rm TeV}$.}
\label{LNV2}
\end{figure}

To see the interference effects indicated by the `hill-valley' structure for {\bf TIII-NP} in the blue curves in Fig.~\ref{LNV1}(a) and in Fig.~\ref{LNV1}(b) more precisely, we plot $\sigma(\mu^\pm \mu^\pm \rightarrow W_i^\pm W_i^\pm), (i=1,2)$ versus $\sqrt s$ in Fig.~\ref{LNV2}, where (a) for $i=1$, (b) for $i=2$, and the solid, dashed, dotted curves in the figures denote the results with $M_{N_2}=1.0,\;2.0,\;3.0\;{\rm TeV}$ respectively. In Fig.~\ref{LNV2}~(a) the black curves denote the results for {\bf TI-NP}, the red curves in Fig.~\ref{LNV2}~(b) denote the results for {\bf TII-NP} with $M_{W_2}=5.0\;{\rm TeV}$, and the blue curves in both the figures denote the results for {\bf TIII-NP} with $M_{W_2}=5.0\;{\rm TeV}$, $Y_{\mu\mu}=1.0$, $M_{\Delta_L^{\pm\pm}}=10.0\;{\rm TeV}$, $M_{\Delta_R^{\pm\pm}}=11.0\;{\rm TeV}$.
The interference effects between the contributions from doubly charged Higgs and those from the Majorana neutral leptons can be seen clearly (the blue curves) in Fig.~\ref{LNV2}, and both $\sigma(\mu^\pm \mu^\pm \rightarrow W_1^\pm W_1^\pm )$, $\sigma(\mu^\pm \mu^\pm \rightarrow W_2^\pm W_2^\pm )$ increase with increasing $M_{N_2}$. The heights of the peaks on the three blue curves in Fig.~\ref{LNV2}~(a) are similar, because the contributions to $\sigma(\mu^\pm \mu^\pm \rightarrow W_1^\pm W_1^\pm )$ are dominated by $\Delta_L^{\pm\pm}$ contributions at the resonance, and varying $M_{N_2}$ does not affect the heights of the resonance peaks at all. However Fig.~\ref{LNV2} (b) shows that the heights of the peaks of the three blue curves vary with $M_{N_2}$, because $M_{N_2}$ is related to the Yukawa coupling $\Delta_R^{\pm\pm} \mu^{\mp}\mu^{\mp}$ and the resonance heights are dominated by the Yukawa coupling.

\begin{figure}
\setlength{\unitlength}{1mm}
\centering
\includegraphics[width=3.5in]{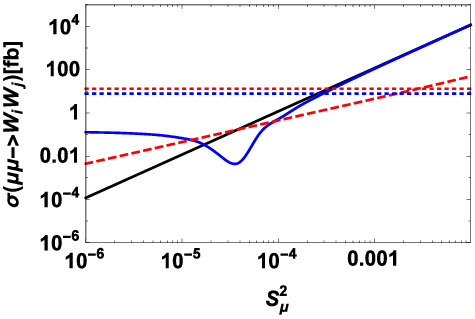}
\vspace{0cm}
\caption[]{$\sigma(\mu^\pm \mu^\pm \rightarrow W_i^\pm W_j^\pm)$ versus $S_\mu^2$ with $M_{N_2}=2.0\;{\rm TeV}$, $\sqrt s=15.0\;$TeV (the rest relevant parameters are set the same as those in Fig.~\ref{LNV2}. The black solid curve denotes the result of $W_iW_j=W_1W_1$ for {\bf TI-NP}. The red dashed, red dotted curves denote the results of $W_iW_j=W_1W_2$, $W_iW_j=W_2W_2$ respectively for {\bf TII-NP} with $M_{W_2}=5.0\;$TeV. The blue solid, blue dotted curves denote the results of $W_iW_j=W_1W_1$, $W_iW_j=W_2W_2$ respectively for {\bf TIII-NP} with $M_{W_2}=5.0\;$TeV, $Y_{\mu\mu}=1.0$, $M_{\Delta_L^{\pm\pm}}=10.0\;$TeV, $M_{\Delta_R^{\pm\pm}}=11.0\;$TeV.}
\label{LNV3}
\end{figure}

To explore and to see the effects of $S_\mu^2$ to the cross-sections $\sigma(\mu^\pm \mu^\pm \rightarrow W_i^\pm W_j^\pm)$, we set $M_{N_2}=2.0\;{\rm TeV}$, $\sqrt s=15.0\;$TeV and the rest relevant parameters as the same as those in Fig.~\ref{LNV2}, and plot $\sigma(\mu^\pm \mu^\pm \rightarrow W_i^\pm W_j^\pm)$ versus $S_\mu^2$ in Fig.~\ref{LNV3}. The black solid curve denotes the result of $W_iW_j=W_1W_1$ for {\bf TI-NP}. The red dashed, red dotted curves denote the results of $W_iW_j=W_1W_2$, $W_iW_j=W_2W_2$ respectively for {\bf TII-NP} with $M_{W_2}=5.0\;$TeV. The blue solid, blue dotted curves denote the results of $W_iW_j=W_1W_1$, $W_iW_j=W_2W_2$ respectively for {\bf TIII-NP} with $M_{W_2}=5.0\;$TeV, $Y_{\mu\mu}=1.0$, $M_{\Delta_L^{\pm\pm}}=10.0\;$TeV, $M_{\Delta_R^{\pm\pm}}=11.0\;$TeV. The dotted curves in Fig.~\ref{LNV3} indicate $\sigma(\mu^\pm \mu^\pm \rightarrow W_2^\pm W_2^\pm )$ is not sensitive to $S_\mu^2$, and this fact can be read out from Eq.~(\ref{33}). The results described by the black solid curve and the red dashed curve show that $\sigma(\mu^\pm \mu^\pm \rightarrow W_1^\pm W_1^\pm )$ for {\bf TI-NP}, {\bf TII-NP} and $\sigma(\mu^\pm \mu^\pm \rightarrow W_1^\pm W_2^\pm )$ for {\bf TII-NP}, {\bf TIII-NP} increase with increasing $S_\mu^2$. As shown by the blue solid curves in the figures, $\sigma(\mu^\pm \mu^\pm \rightarrow W_1^\pm W_1^\pm )$ for {\bf TIII-NP} decreases with $S_\mu^2$ increasing first and then increases with $S_\mu^2$ increasing. It is due to the fact that the contributions to the process $\mu^\pm \mu^\pm \to W_1^\pm W_1^\pm$ for {\bf TIII-NP} are dominated by the doubly charged Higgs when $S_\mu^2$ is small ($S_\mu^2\lesssim 10^{-5}$), while the interference effects of the contributions from Majorana neutral leptons and from the doubly charged Higgs become important when $S_\mu^2\approx 10^{-4}$ (the interference effects can be seen more clearly in Fig.~\ref{LNV2}), and the Majorana neutral lepton contributions play dominant roles when $S_\mu^2$ is large enough ($S_\mu^2\gtrsim 10^{-3}$).

It is pointed in Ref.~\cite{Li:2023tbx} that a high energy $\mu^+\mu^-$ collider running at the collision energy $\sqrt s\simeq 10\;{\rm TeV}$ can probe $S_\mu^2$ down to the region $\mathcal{O} (10^{-7}) \sim \mathcal{O}(10^{-4})$ if a integrated luminosity $10^4\;{\rm fb}^{-1}$ is accumulated. At the same sign high energy muon colliders with a collision energy such as $\sqrt s\simeq 15\;{\rm TeV}$, the solid black curve in Fig.~\ref{LNV3} shows that the total cross section of the LNV process $\mu^-\mu^-\rightarrow W_1^- W_1^-$ may reach up-to about $0.01\;{\rm fb}$ by the prediction of {\bf TI-NP} with $S_\mu^2\approx 10^{-5}$. Note from the blue and red curves in Fig.~\ref{LNV3} that the cross sections of the LNV processes $\mu^\pm\mu^\pm\rightarrow W_1^\pm W_1^\pm$ predicted by {\bf TII-NP} and {\bf TIII-NP} are larger than the one of predicted by {\bf TI-NP} when $S_\mu^2 < 10^{-4}$, hence the LNV processes predicted by {\bf TII-NP} and {\bf TIII-NP} at a high energy $\mu^\pm\mu^\pm$ collider have much greater opportunities to be observed. The observation of these processes must offers unambiguous evidences of NP, which is a totally different way to fix the parameter $S_\mu^2$ from that at the high energy $\mu^+\mu^-$ colliders~\cite{Li:2023tbx}. Moreover the same-sign muon colliders have very special advantages in observing the doubly charged Higgs, and setting constraints on the couplings of doubly charged Higgs with charged leptons etc. From the facts mentioned here particularly, the complementarity of the high energy same-sign muon colliders with the high energy $\mu^+\mu^-$ colliders can be concluded very well.

\begin{figure}
\setlength{\unitlength}{1mm}
\centering
\includegraphics[width=1.9in]{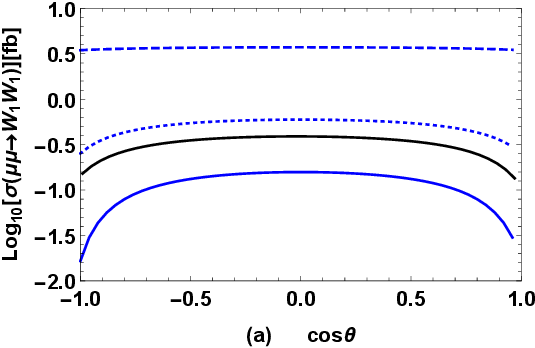}
\vspace{0.1cm}
\includegraphics[width=1.9in]{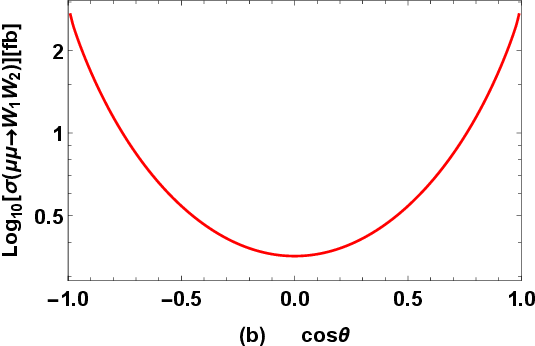}
\vspace{0.1cm}
\includegraphics[width=1.9in]{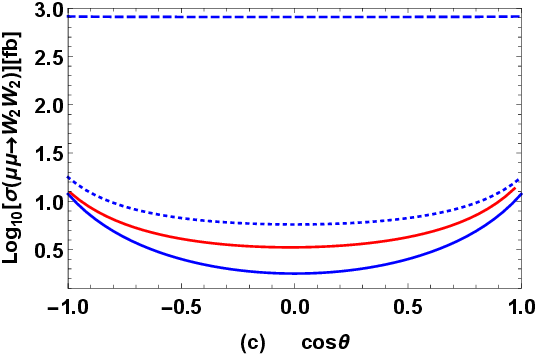}
\vspace{0cm}
\caption[]{Taking $M_{N_2}=2\;{\rm TeV}$, $S_\mu^2=10^{-4}$, (a): the angle distributions of $\mu^\pm \mu^\pm \rightarrow W_1^\pm W_1^\pm $ for $\sqrt s=5\;{\rm TeV}$, where the black curve denotes the results for {\bf TI-NP}, the blue curves denote the results for {\bf TIII-NP} with $M_{W_2}=5\;{\rm TeV}$, $Y_{\mu\mu}=1$ and the blue solid, blue dashed, blue dotted curves denote the results for $M_{\Delta^{\pm\pm}_L}=M_{\Delta^{\pm\pm}_R}=3,\;5,\;7\;$TeV respectively; (b): the angle distributions of $\mu^\pm \mu^\pm \rightarrow W_1^\pm W_2^\pm $ for $\sqrt s=7\;$TeV, where the red curve denotes the results for {\bf TII-NP}; (c): the angle distributions of $\mu^\pm \mu^\pm \rightarrow W_2^\pm W_2^\pm $ for $\sqrt s=12\;$TeV, where the red curve denotes the results for {\bf TII-NP}, the blue curves denote the results for {\bf TIII-NP} with $Y_{\mu\mu}=1$, and the blue solid, blue dashed, blue dotted curves denoting the results for $M_{\Delta^{\pm\pm}_L}=M_{\Delta^{\pm\pm}_R}=7,\;12,\;15\;$TeV respectively.}
\label{LNV4}
\end{figure}

Taking $M_{N_2}=2\;{\rm TeV}$, $S_\mu^2=10^{-4}$, we plot the angle distributions of $\mu^\pm \mu^\pm \rightarrow W_i^\pm W_j^\pm $ in Fig.~\ref{LNV4}. Fig.~\ref{LNV4} (a) is that of $W_iW_j=W_1W_1$ with $\sqrt s=5\;{\rm TeV}$, where the black curve denotes the results for {\bf TI-NP}, the blue curves denote the results for {\bf TIII-NP} with $M_{W_2}=5\;{\rm TeV}$, $Y_{\mu\mu}=1$ and the blue solid, blue dashed, blue dotted curves denote the results with $M_{\Delta^{\pm\pm}_L}=M_{\Delta^{\pm\pm}_R}=3,\;5,\;7\;$TeV respectively. Fig.~\ref{LNV4} (b) is that of $W_iW_j=W_1W_2$ with $\sqrt s=7\;{\rm TeV}$, where the red curve denotes the results obtained for {\bf TII-NP}. Fig.~\ref{LNV4} (c) is that of $W_iW_j=W_2W_2$ with $\sqrt s=12\;{\rm TeV}$, where the red curve denotes the results for {\bf TII-NP}, the blue curves denote the results for {\bf TIII-NP} with $Y_{\mu\mu}=1$, and the blue solid, blue dashed, blue dotted curves denote the results for $M_{\Delta^{\pm\pm}_L}=M_{\Delta^{\pm\pm}_R}=7,\;12,\;15\;$TeV respectively.

As shown in the figures, the angular distributions of the process $\mu^\pm \mu^\pm \rightarrow W_i^\pm W_i^\pm\;(i=1,2)$ predicted by {\bf TIII-NP} are flat for $\sqrt s\approx M_{\Delta^{\pm\pm}_L}=M_{\Delta^{\pm\pm}_R}$, because the contributions to the processes are dominated by the s-channel doubly charged Higgs in this case. The black solid, blue solid and blue dotted curves in Fig.~\ref{LNV4} (a) show that $\sigma(\mu\mu\rightarrow W_1W_1)$ takes the maximum value at $\cos\theta=0$, and the contributions to the process are dominated by Majorana neutral leptons in these three cases. Fig.~\ref{LNV4} (b), (c) show that $\sigma(\mu\mu\rightarrow W_1W_2)$ predicted by {\bf TII-NP}, {\bf TIII-NP} and $\sigma(\mu\mu\rightarrow W_2W_2)$ predicted by {\bf TII-NP} take the minimum values at $\cos\theta=0$. The angular distributions of the process $\mu^\pm \mu^\pm \rightarrow W_2^\pm W_2^\pm$ predicted by {\bf TIII-NP} are flat when the contributions are dominated by doubly charged Higgs, and $\sigma(\mu\mu\rightarrow W_2W_2)$ takes the minimum value at $\cos\theta=0$ when the contributions are dominated by Majorana neutral leptons.

\section{Discussions and Summary\label{sec5}}

If it is feasible to build a high-energy $\mu^+\mu^-$ collider in future when the progresses on the relevant techniques are achieved, then to build a very high energy same-sign $\mu^\pm \mu^\pm$ collider should have no additional serious technical problems. Therefore, it is crucial to investigate important physics phenomena, such as lepton di-flavor violation (LFV) processes $\mu^\pm \mu^\pm \rightarrow e^\pm e^\pm $, $\mu^\pm \mu^\pm \rightarrow \tau^\pm \tau^\pm $ and the lepton di-number violation (LNV) processes $\mu^\pm \mu^\pm \rightarrow W_i^\pm W_j^\pm$ ($i,\;j=1,\;2$) etc, at very high energy $\mu^\pm \mu^\pm $ colliders. All of these processes are forbidden in the SM, the observable of these processes is sensitive to the nature of the heavy neutral leptons and neutrinos, e.g. the LFV and LNV physics at TeV energy scale and the doubly charged Higgs in certain new physics (NP) models. In this regard, we have performed quantitative evaluations of the NP contributions to these processes and explored their essential characteristics by categorizing the involved NP factors into three types. Taking into account the constraints imposed by existing experiments, we have computed and appropriately presented the numerical results in figures. This section will provide brief discussions of the obtained results and summarizes the significance of observing LFV di-lepton and LNV di-boson processes at TeV-scale $\mu^\pm \mu^\pm$ colliders.

For the LFV di-lepton processes, Figs.~\ref{LFV1}-\ref{LFV3} clearly show that the predicted $\sigma(\mu^\pm \mu^\pm \rightarrow \tau^\pm \tau^\pm )$ and $\sigma(\mu^\pm \mu^\pm \rightarrow e^\pm e^\pm )$ can reach to $10\;{\rm fb}$ and $10^{-6}\;{\rm fb}$ respectively for {\bf TI-NP} and {\bf TII-NP}. It indicates that the process $\mu^\pm \mu^\pm \rightarrow \tau^\pm \tau^\pm$ has great opportunities to be observed at the high energy $\mu^\pm \mu^\pm $ colliders, while the process $\mu^\pm \mu^\pm \rightarrow e^\pm e^\pm$ is comparatively hard to be observed at such $\mu^\pm \mu^\pm$ colliders. For {\bf TIII-NP}, the contributions to the LFV di-lepton processes are dominated by doubly charged Higgs in most cases, and the results are much larger than the ones for {\bf TI-NP} and {\bf TII-NP}. Both the processes $\mu^\pm \mu^\pm \rightarrow \tau^\pm \tau^\pm$ and $\mu^\pm \mu^\pm \rightarrow e^\pm e^\pm$ predicted by {\bf TIII-NP} have great opportunities to be observed at high energy $\mu^\pm \mu^\pm $ colliders. Furthermore, Fig.~\ref{LFV2} clearly demonstrates the resonance enhancements caused by the two $s-$channel doubly charged Higgs for {\bf TIII-NP}. This implies that the LFV processes $\mu^\pm\mu^\pm\rightarrow \tau^\pm \tau^\pm$ and $\mu^\pm\mu^\pm\rightarrow e^\pm e^\pm$ serve as favorable channels for observing the two doubly charged Higgs at high-energy $\mu^\pm \mu^\pm$ colliders. The angular distributions of $\mu^\pm \mu^\pm \rightarrow \tau^\pm \tau^\pm $ and $\mu^\pm \mu^\pm \rightarrow e^\pm e^\pm $ predicted for {\bf TI-NP}, {\bf TII-NP} and {\bf TIII-NP} are flat which indicates that the di-lepton processes predicted in all these three types of NP are insensitive to the angular cuts (Fig.~\ref{LFV4}).

For the LNV di-boson processes, the results for $\sigma(\mu^\pm\mu^\pm\rightarrow W_1^\pm W_1^\pm)$ for {\bf TII-NP} are similar to the ones for $\sigma(\mu^\pm\mu^\pm\rightarrow W_L^\pm W_L^\pm)$ for {\bf TI-NP}, and the results for $\sigma(\mu^\pm\mu^\pm\rightarrow W_1^\pm W_2^\pm)$ for {\bf TIII-NP} are similar to the ones for $\sigma(\mu^\pm\mu^\pm\rightarrow W_1^\pm W_2^\pm)$ for {\bf TII-NP}. Figs.~\ref{LNV1}-\ref{LNV3} clearly demonstrate that $\sigma(\mu^\pm \mu^\pm \rightarrow W_1^\pm W_1^\pm )$ predicted by {\bf TI-NP} and {\bf TII-NP} can reach a large value of about $10^4\;{\rm fb}$ when $\sqrt s=15\;{\rm TeV}$, the cross-section $\sigma(\mu^\pm \mu^\pm \rightarrow W_1^\pm W_2^\pm )$ predicted by {\bf TII-NP} and {\bf TIII-NP} can reach to about $100\;{\rm fb}$ when $\sqrt s=15\;{\rm TeV}$. The resonance enhancements in $\sigma(\mu^\pm \mu^\pm \rightarrow W_1^\pm W_1^\pm )$ and $\sigma(\mu^\pm \mu^\pm \rightarrow W_2^\pm W_2^\pm )$ are due to the contributions from doubly charged Higgs for {\bf TIII-NP}. Thus, observing the processes $\mu^\pm \mu^\pm \rightarrow W_1^\pm W_1^\pm$, $\mu^\pm \mu^\pm \rightarrow W_2^\pm W_2^\pm$ at very high energy $\mu^\pm\mu^\pm$ colliders represents a valuable approach to the signals originated either from $\Delta_L^{\pm\pm}$ or from $\Delta_R^{\pm\pm}$ when observing the relevant signals. Measuring the total decay widths of $\Delta_L^{\pm\pm}$ and $\Delta_R^{\pm\pm}$ for {\bf TIII-NP} via observing the LNV di-boson processes is also an important topic for the high energy $\mu^\pm \mu^\pm$ colliders. The results of angle distributions (Fig.~\ref{LNV4}) show that ${\rm d}\sigma(\mu\mu\rightarrow W_1W_1)/{\rm d}\cos\theta$ predicted for {\bf TI-NP}, {\bf TII-NP} takes the maximum value at $\cos\theta=0$. For {\bf TIII-NP}, the angular distributions of the process $\mu^\pm \mu^\pm \rightarrow W_1^\pm W_1^\pm$ are flat when the contributions to $\mu^\pm \mu^\pm \rightarrow W_1^\pm W_1^\pm$ are dominated by doubly charged Higgs, and $\sigma(\mu\mu\rightarrow W_1W_1)$ takes the maximum value at $\cos\theta=0$ when the contributions are dominated by Majorana neutral leptons. $\sigma(\mu\mu\rightarrow W_1W_2)$ predicted for {\bf TII-NP}, {\bf TIII-NP} and $\sigma(\mu\mu\rightarrow W_2W_2)$ predicted for {\bf TII-NP} take the minimum values at $\cos\theta=0$. The angular distributions of the process $\mu^\pm \mu^\pm \rightarrow W_2^\pm W_2^\pm$ predicted for {\bf TIII-NP} are flat when the contributions are dominated by doubly charged Higgs, and $\sigma(\mu\mu\rightarrow W_2W_2)$ takes the minimum value at $\cos\theta=0$ when the contributions are dominated by Majorana neutral leptons.

In summary, observing the leptonic di-flavor violation (LFV) and di-number violation (LNV) processes represents one of the most important physics aspects at high-energy $\mu^\pm\mu^\pm $ colliders. The quantitative investigations in this paper lead us to conclude that the LFV process $\mu^\pm \mu^\pm \rightarrow \tau^\pm \tau^\pm$ predicted by {\bf TI-NP}, {\bf TII-NP} and {\bf TIII-NP} is highly expected to be observed at very high-energy $\mu^\pm\mu^\pm$ colliders, while the process $\mu^\pm \mu^\pm \rightarrow e^\pm e^\pm$ is expected to be observed only for {\bf TIII-NP}. And all of the LNV di-boson processes predicted by {\bf TI-NP}, {\bf TII-NP} and {\bf TIII-NP} have great opportunities to be observed at such colliders. Based on the numerical results analyzed in Sec. ~\ref{sec4}, one can have more insights into the characteristics of the processes due to NP: {\bf TI}, {\bf TII}, {\bf TIII} respectively, and will achieve relevant constraints on parameters such as $M_N$, $M_{W_2}$, $Y_{\mu\mu}$, $M_{\Delta_L^{\pm\pm}}$, $M_{\Delta_R^{\pm\pm}}$. It should be emphasized that at such high energy $\mu^\pm\mu^\pm$ collider, there are significant opportunities to observe the NP, such as the leptonic Majorana components, right-handed $W$-bosons and the two doubly charged Higgs and their properties etc.

Therefore, owing to the fact that there are so many important physics and no serious problems in building high energy same-sign muon colliders as explored in this paper we believe that in the future high-energy $\mu^+\mu^-$ colliders and same-sign $\mu^\pm\mu^\pm$ colliders both will be built when the techniques on muon acceleration, muon beam storage in a circular ring and collisions of two counter-propagating muon beams etc are matured.

\vspace{5mm}

\noindent {\bf\Large Acknowledgments:} This work was supported in part by the National Natural Science
Foundation of China (NNSFC) under Grants No. 12047503, No. 12075301, No. 11821505£¬
and No. 11705045. It was also supported in part by the Key Research Program
of Frontier Sciences, CAS, Grant No. QYZDY-SSW-SYS006. The authors
(J-L Yang and C-H Chang) would like to thank Prof S-S Bao (SDU) for helpful discussions about MadGraph5.

\end{document}